\documentclass[prd,a4paper,showpacs,
nofootinbib
]{revtex4}
\usepackage{graphicx}
\usepackage{epsfig}
\def\eq#1{{Eq.~(\ref{#1})}}

\newcommand{\be}{\begin{equation}}
\newcommand{\ee}{\end{equation}}
\newcommand{\bea}{\begin{eqnarray}}
\newcommand{\eea}{\end{eqnarray}}

\newcommand{\ms}{\Delta m^2_{\odot}}
\newcommand{\ma}{\Delta m^2_{\rm atm}}

\newcommand{\sss}{\sin^2 \theta_{\odot}}
\newcommand{\sch}{\sin^2 \theta}
\newcommand{\thsol}{\mbox{$\theta_{\odot}$~}}
\newcommand{\kl}{\mbox{KamLAND~}}
\newcommand{\beq}{\begin{equation}}
\newcommand{\eeq}{\end{equation}}

\def\ltap{\ \raisebox{-.4ex}{\rlap{$\sim$}} \raisebox{.4ex}{$<$}\ }
\def\gtap{\ \raisebox{-.4ex}{\rlap{$\sim$}} \raisebox{.4ex}{$>$}\ }
\newcommand{\deltaatm}{\mbox{$\Delta m^2_{\mathrm{atm}}$}}
\newcommand{\deltasol}{\mbox{$ \Delta m^2_{\odot}$}}
\def\gtap{\mathrel{ \rlap{\raise 0.511ex \hbox{$>$}}{\lower 0.511ex
   \hbox{$\sim$}}}} 
\def\ltap{\mathrel{ \rlap{\raise 0.511ex
   \hbox{$<$}}{\lower 0.511ex \hbox{$\sim$}}}}
\begin{document}

\begin{flushright}
SISSA 48/2003/EP\\
hep-ph/0306017
\end{flushright}

\title{Precision Neutrino Oscillation Physics with an 
Intermediate Baseline Reactor Neutrino 
Experiment}

\author{Sandhya Choubey$^{1,2}$}
\author{S.T. Petcov$^{2,1,3}$}
\author{M. Piai$^4$}
\affiliation{$^1$INFN, Sezione di Trieste, Trieste, Italy}
\affiliation{$^2$Scuola Internazionale Superiore di Studi Avanzati, 
I-34014 Trieste, Italy}
\affiliation{$^3$Institute of Nuclear Research and
Nuclear Energy, Bulgarian Academy of Sciences, 1784 Sofia, Bulgaria}
\affiliation{$^4$Department of Physics, Yale University, New Haven CT 06520 USA}

\begin{abstract}
We discuss the physics potential
of intermediate $L \sim 20 \div 30$ km baseline
experiments at reactor facilities,
assuming that the solar 
neutrino oscillation 
parameters $\ms$ and \thsol
lie in the high-LMA solution region.
We show that such an intermediate 
baseline reactor experiment can 
determine both $\ms$ and 
\thsol with a remarkably high precision.
We perform also a detailed study of 
the sensitivity of the indicated experiment 
to $\Delta m^2_{\rm atm}$, which drives
the dominant atmospheric 
$\nu_{\mu}$ ($\bar{\nu}_{\mu}$)
oscillations, and to $\theta$ - the
neutrino mixing angle limited by the data from
the CHOOZ and Palo Verde experiments.
We find that this experiment can 
improve the bounds on $\sch$. 
If the value of $\sch$ is large enough, 
$\sin^2\theta \gtap 0.02$,  
the energy resolution 
of the detector is sufficiently good 
and if the statistics is relatively high,
it can determine with extremely high precision 
the value of $\Delta m^2_{\rm atm}$.
We also explore the potential of the
intermediate baseline reactor neutrino experiment  
for determining the type of the 
neutrino mass spectrum, 
which can be with normal 
or inverted hierarchy. We  show that the 
conditions under which the type of neutrino 
mass hierarchy can be determined are 
quite challenging, but are within the 
reach of the experiment under discussion.

\end{abstract}

\preprint{SISSA ??/03/EP}
\preprint{ YCTP-XX-03}

\maketitle


\section{Introduction}
\label{section:introduction}


 The experiments with solar, atmospheric and reactor neutrinos
~\cite{sol, SKsolar, SNO, SKatm, K2K, KamLAND}
have provided in recent years remarkable 
evidences for the existence of neutrino oscillations 
driven by nonzero neutrino masses and neutrino mixing.
The hypothesis of solar neutrino oscillations, 
which in one variety or another 
were considered as 
the most natural explanation 
of the solar neutrino deficit~\cite{sol,SKsolar}
since the late 60'ies 
(see, e.g., \cite{BPont67, BiPet87, SPSchlad97}),
has received a convincing
confirmation from the measurement 
of the solar neutrino flux
through the neutral current reaction on
deuterium by the 
SNO experiment \cite{SNO}. The 
analysis of the solar neutrino data
obtained by 
Homestake, SAGE, GALLEX/GNO, SK and SNO
experiments showed that the data favor 
the Large Mixing Angle (LMA) MSW 
solution with the two-neutrino 
oscillation parameters -
the solar neutrino mixing angle and the mass
squared difference,
lying at $99.73\%$ C.L. in the region \cite{Choubey:2002nc, SNO}:
\bea
 3 \times 10^{-5} {\mbox{eV}}^2 & \ltap \ms
\ltap &  3.5 \times 10^{-4} {\mbox{eV}}^2 
\label{sollimit}
\\
0.21 & \ltap \sin^2\theta_{\odot} \ltap & 0.47\,. 
\label{thetalimit}
\eea

\noindent  
\indent The first results of the 
KamLAND reactor experiment
\cite{KamLAND} has confirmed,
under the plausible assumption of CPT-invariance
which we will suppose to hold throughout this study,
the LMA MSW solution, establishing it essentially as 
a unique solution of the solar neutrino problem. 
The combined fits to the available
solar neutrino and KamLAND  
data, performed by several collaborations
within the two-neutrino mixing hypothesis,
identify two distinct solution sub-regions within 
the LMA solution region~\cite{SolFit1, SolFit2, SolFit3}.
Adding the KamLAND data 
did not lead to a considerable reduction of the
interval of allowed values of $\sss$
with respect to the one quoted in eq. (\ref{sollimit}),
while the best fit values of 
$\ms$ in the two sub-regions (labeled from now on 
low-LMA and high-LMA) are given by \cite{SolFit2}:
\bea \mbox{low-LMA:}\,&\,\ms\,\simeq\,&\,7.2\,\times\,10^{-5}\,
\mbox{eV}^2\,,\\
\mbox{high-LMA:}\,&\,\ms\,\simeq\,&\,1.5\,\times\,10^{-4}\,
\mbox{eV}^2\,. 
\label{postKamL}
\eea

%
\indent  The observed Zenith angle dependence of 
the multi-GeV $\mu-$like events in the Super-Kamiokande
experiment unambiguously demonstrated
the disappearance of the atmospheric  
$\nu_{\mu}$ ($\bar{\nu}_{\mu}$) 
on distances $L \gtap 1000$ km.
The Super-Kamiokande (SK) atmospheric neutrino data
is best described, as is well-known,
in terms of dominant $\nu_{\mu} \rightarrow \nu_{\tau}$ 
($\bar{\nu}_{\mu}\rightarrow \bar{\nu}_{\tau}$) oscillations
with (almost) maximal mixing and neutrino mass squared difference 
of  $|\Delta m^2_{\rm atm}| \cong (1.4 - 5.0)\times 10^{-3}$ eV$^2$ 
($99.73\%$ C.L.) ~\cite{SKatm}.
According to the more recent 
combined analysis of the data from 
the SK and K2K experiments
\cite{Fogli:2003th} one has 
at 99.73\% C.L.:
\bea 
1.5 \,\times\, 10^{-3}\,\mbox{eV}^2\,\ltap\,|\Delta
m^2_{\mbox{atm}}|\, \ltap\,3.7\, \times\,10^{-3}\,\mbox{eV}^2 
\eea
%

 The neutrino oscillation 
description of the combined solar and 
atmospheric neutrino data requires
the existence of three-neutrino mixing 
in the weak charged lepton current
(see, e.g., \cite{BGG99}):
\begin{equation}
\nu_{l \mathrm{L}}  = \sum_{j=1}^{3} U_{l j} \, \nu_{j \mathrm{L}}~.
\label{3numix}
\end{equation}
\noindent Here $\nu_{lL}$, $l  = e,\mu,\tau$,
are the three left-handed flavor 
neutrino fields,
$\nu_{j \mathrm{L}}$ is the 
left-handed field of the 
neutrino $\nu_j$ having a mass $m_j$
and $U$ is the Pontecorvo-Maki-Nakagawa-Sakata (PMNS)
neutrino mixing matrix \cite{BPont57, MNS62}.
The PMNS mixing matrix U can be parameterized by 
three angles, $\theta_{\rm atm}$,
$\theta_{\odot}$, and $\theta$,
and, depending on whether the massive 
neutrinos $\nu_j$ are Dirac or Majorana particles -
by one or three CP-violating phases \cite{BHP80,Doi81}.
In the standard parameterization 
of U (see, e.g., \cite{BGG99}) 
the three mixing angles are denoted as
$\theta_{12}$, $\theta_{13}$ and
$\theta_{23}$:
\begin{equation}
U = \left(\begin{array}{ccc}
U_{e1}& U_{e2} & U_{e3} \\
U_{\mu 1} & U_{\mu 2} & U_{\mu 3} \\
U_{\tau 1} & U_{\tau 2} & U_{\tau 3} 
\end{array} \right)
= \left(\begin{array}{ccc} 
c_{12}c_{13} & s_{12}c_{13} & s_{13}e^{-i\delta}\\
 - s_{12}c_{23} - c_{12}s_{23}s_{13}e^{i\delta} & 
c_{12}c_{23} - s_{12}s_{23}s_{13}e^{i\delta} & s_{23}c_{13}\\
s_{12}s_{23} - c_{12}c_{23}s_{13}e^{i\delta} 
& -c_{12}s_{23} - s_{12}c_{23}s_{13}e^{i\delta}
& c_{23}c_{13}\\ 
\end{array} \right)
\label{Umix}
\end{equation}
%

%
\noindent where we have used the usual notations, 
$s_{ij} \equiv \sin \theta_{ij}$,
$c_{ij} \equiv \cos \theta_{ij}$, 
and $\delta$ is the Dirac CP-violation phase
\footnote{We have not 
written explicitly the two possible Majorana CP-violation phases 
\cite{BHP80,Doi81}
which do not enter into the expressions for the oscillation 
probabilities of interest \cite{BHP80,Lang86}.  
We assume throughout this study $0 \leq \theta_{12}, \theta_{23}, 
\theta_{13} < \pi/2$. 
}.
If we identify 
the two independent neutrino mass squared 
differences in this case,
$\Delta m^2_{21}$ and $\Delta m^2_{31}$,  
with the neutrino mass squared differences
which drive the solar and atmospheric 
neutrino oscillations, $\deltasol = \Delta m^2_{21} > 0$,
$\deltaatm = \Delta m^2_{31}$,
one has: $\theta_{12} = \theta_{\odot}$, 
$\theta_{23} = \theta_{\rm atm}$, 
and $\theta_{13} = \theta$. 

  The angle $\theta$ is limited by 
the data from the CHOOZ and Palo Verde
experiments~\cite{CHOOZ,PaloV}
which searched for evidences 
for oscillations of reactor $\bar{\nu}_e$
at $\sim 1$ km from the source. 
No disappearance of $\bar{\nu}_e$ was observed.
In a two-neutrino oscillation
analysis performed in \cite{CHOOZ}
a stringent  upper bound on the value of $\theta$ 
in the region of  $|\Delta m^2| \geq
1.5 \times 10^{-3} \mathrm{eV}^2$ was obtained. 
A 3-$\nu$ oscillation analysis of the CHOOZ data
\footnote{In this case  $\Delta m^2 = 
\Delta m^2_{\rm atm}$.}
\cite{BNPChooz} 
led to the conclusion that 
for $\deltasol  \ltap 10^{-4}~{\rm eV^2}$,
the limits on $\sin^2\theta$ 
practically coincide with
those derived in the 2-$\nu$ 
oscillation analysis in \cite{CHOOZ}.
A combined 3-$\nu$ oscillation
analysis of the 
solar neutrino, CHOOZ and the 
KamLAND data, performed
under the assumption of 
$\deltasol \ll |\deltaatm|$
(see, e.g., \cite{BGG99,ADE80}),
showed that \cite{SolFit1} 
\begin{equation}
\sin^{2} \theta < 0.05,~~~~ 99.73\%~{\rm C.L.}
\label{chooz1}
\end{equation}

%
\noindent The precise upper limit in \eq{chooz1} is 
$\deltaatm$-dependent.
The authors of  
\cite{SolFit1} found
the best-fit value of $\sin^2\theta$
to lie in the interval 
\footnote{The possibility of large 
$\sin^2 \theta > 0.97$, which is admitted 
by the CHOOZ data alone, is incompatible
with the neutrino oscillation interpretation of the solar
neutrino data (see, e.g., \cite{solvsreactor}).}
$\sin^2\theta \cong (0.00 - 0.01)$.

  Somewhat better limits on $\sin^2 \theta$ than 
the existing one can be obtained in the 
MINOS experiment \cite{MINOS}. 
Various options are being currently discussed
(experiments with off-axis neutrino beams, more precise
reactor antineutrino and long baseline experiments, etc.,
see, e.g., \cite{MSpironu02}) of how to improve
by at least an order of magnitude, i.e., 
to values of $\sim 0.005$ or smaller, 
the sensitivity to $\sin^2\theta$. 

   Let us note  that the 
atmospheric neutrino and K2K data
do not allow one to determine the sign
of $\Delta m^2_{\rm atm}$.
This implies that if we identify
$\Delta m^2_{\rm atm}$ with
$\Delta m^2_{31}$
in the case of 3-neutrino mixing, 
one can have $\Delta m^2_{31} > 0$
or $\Delta m^2_{31} < 0$. The two 
possibilities correspond to two different
types of neutrino mass spectrum:
with normal hierarchy (NH), $m_1 < m_2 < m_3$, and 
with inverted hierarchy (IH),
$m_3 < m_1 < m_2$.  

   After the spectacular experimental 
progress made in the last two years or so 
in the studies of 
neutrino oscillations, 
further understanding, in particular,
of the structure of the neutrino masses 
and mixing, of their origins  
and of the status of the CP-symmetry in 
the lepton sector requires a large and
challenging program of high 
precision measurements 
to be pursued in neutrino physics. 
One of the first goals of this 
program is to improve the
precision in the measurement of 
the mass squared differences and
mixing angles which control the 
solar and the dominant atmospheric  
neutrino oscillations: 
$\ms$, $\theta_{\odot}$, 
and $\Delta m^2_{\rm atm}$, $\theta_{atm}$.
A step of fundamental importance 
would be the detection and the studies
of sub-leading neutrino oscillation effects, 
if the latter are observable.
This includes the
measurement of, or getting 
more stringent upper limit on,
the value of the third and 
the only small mixing angle 
in the PMNS matrix $U$, 
$\theta~(=\theta_{13})$,
the exploration of the 
possible CP-violating effects 
and the determination of the type 
of the neutrino mass spectrum which can be with
normal hierarchy (NH) or inverted hierarchy (IH).
Among the further fundamental open questions, 
which cannot be answered by studying neutrino
oscillations, but the progress in the 
studies of which requires a
precise knowledge of the neutrino oscillation 
parameters, are i) the nature ---Dirac or Majorana--- of 
the massive neutrinos, ii) the absolute scale of 
neutrino masses, iii) the mechanism giving rise 
to the neutrino masses and mixing, iv) the possible relation between 
CP-violation in the lepton sector at low energies and
the generation of the baryon asymmetry of the Universe
by the leptogenesis mechanism,
v) lepton number violation and its possible
manifestation in charged 
lepton flavor violating decays, to name a few.

  In the present article we explore the 
possibility of performing a high precision
measurement of some of the 3-neutrino 
oscillation parameters in an
intermediate baseline, $L \sim 20 \div 30$ km,
reactor neutrino experiment. We first address the 
issue of precision determination 
of the solar neutrino  oscillation
parameters, $\ms$ and $\theta_\odot$. 
The \kl experiment, as discussed by several authors 
\cite{KlStudies}, has a remarkable 
sensitivity to $\ms$ 
in the low-LMA region.
However, the sensitivity of 
\kl to the value of \thsol is not found to be 
equally good. Even under the most optimistic 
conditions \kl is not expected to substantially 
reduce the region of allowed values of
\thsol obtained from the
analysis of  the solar neutrino 
data \cite{Ch}. The best conditions for precise 
determination of \thsol is in a reactor 
experimental set-up, where the baseline is tuned to a 
Survival Probability MINimum (SPMIN)
\footnote{A detailed discussion of 
SPMAX and SPMIN as well as \thsol 
sensitivity of the current and
future experiments can be found in \cite{Ch} 
and we do not repeat it here.}.
With a baseline of about $L\sim 160$ km and
$\ms \sim 7.2\times 10^{-5}$ eV$^2$
or $1.5\times 10^{-4}$ eV$^2$,
KamLAND is essentially sensitive to a 
Survival Probability MAXimum (SPMAX). 
The sensitivity to 
SPMAX gives \kl the ability to 
determine $\ms$ with a high precision 
through the measurement of the  
distortion of the final state $e^{+}-$spectrum,
provided $\ms < 2\times 10^{-4}$ eV$^2$. 
At the same time it reduces 
KamLAND's sensitivity 
to the solar neutrino mixing angle \cite{Ch}. 
As was shown in \cite{Ch}, a 
70 km baseline reactor 
experiment could determine the 
solar neutrino mixing angle $\sin^2\theta_{\odot}$
 $(\tan^2\theta_\odot)$ 
to within 9.6 (14)\% at the 99\% C.L. 
in the case of the low-LMA solution
of the solar neutrino problem.
\kl essentially  looses  sensitivity 
to $\ms$ for $\ms \gtap 2\times 10^{-4}$ 
eV$^2$. Nevertheless, the more precise measurement 
of the spectrum of the final state $e^+$
in the \kl experiment is expected
to unambiguously determine whether
$\ms$ lies in the low-LMA or high-LMA region.

   In this paper, following a previous suggestion and
analysis in \cite{SPMPiai01} (see also \cite{HLMA}),
in the context of three neutrino 
mixing, we discuss the physics potential
of intermediate $L \sim 20 \div 30$ km baseline
experiments at reactor facilities. 
We show that such an intermediate 
baseline reactor experiment can 
determine both $\ms$ and 
\thsol with a remarkably high precision
{\it if the solution of 
the solar neutrino problem is
the high-LMA solution}. 
We perform also a detailed study of 
the sensitivity of the indicated experiment 
to the parameters 
$\Delta m^2_{\rm atm}$ and $\sch$.
We show that if the energy resolution 
of the detector is sufficiently good 
and if the statistics is relatively high, 
one can choose small enough energy bins 
so that even an experiment with an 
intermediate baseline of $20-30$ km can be sensitive to 
the $\Delta m^2_{\rm atm}$ driven oscillations.
We find that this experiment can 
certainly improve the bounds on 
$\sch$ and if the value of $\sch$ is large enough, 
it can determine with extremely high precision 
the value of $\Delta m^2_{\rm atm}$.

   We finally explore the potential of the
intermediate baseline reactor neutrino experiment  
for determining the type of the 
neutrino mass spectrum 
which can be with normal or inverted hierarchy. 
The knowledge of the neutrino mass hierarchy
is of crucial importance, in particular,
for modeling of the neutrino mass matrix and 
understanding of the underlying physics
of neutrino mass generation.
The type of the neutrino mass hierarchy 
has been also shown to be an 
important ``parameter'' 
in a number of neutrino mixing and 
neutrino oscillation observables,
such as the effective Majorana mass 
in neutrinoless double-$\beta$
decay \cite{BPP1}.
The neutrino mass hierarchy 
(or, e.g., the sign of $\ma$)
can be determined in
very long baseline neutrino oscillation 
experiments at neutrino factories
(see, e.g., \cite{AMMS99}), and, e.g, using
combined data from the long baseline
oscillation experiments JHF-SK and
NuMI with off-axis neutrino beams \cite{HLM}.
It was suggested in \cite{SPMPiai01} 
that the ``interference'' effects 
between the $\ms$ and $\deltaatm$ driven 
oscillations can be used in reactor 
experiments to answer the question
about the neutrino mass hierarchy.
We show that the 
conditions under which 
the type of neutrino 
mass hierarchy can be 
determined are rather
challenging, but are 
are not out of
reach of the experiment under discussion, 
provided the solar neutrino 
oscillation parameters lie in the 
high-LMA solution region.


\section{The three-generation $\bar{\nu}_e$ survival probability}
\label{section:formulae}

 The expression for the
$\bar{\nu}_e$ survival probability in the case
of 3 flavor neutrino mixing and neutrino mass spectrum 
with normal hierarchy (NH) 
is given by 
\footnote{The Earth matter effects are
negligible for the values of the neutrino oscillation 
parameters ($\ms$ and $\deltaatm$),
$\bar{\nu}_e$ energies and 
the short baseline $L\simeq 20 \div 30$ km we are
interested in.}
\cite{BNPChooz, SPMPiai01, SChReactP}: 

\bea
\lefteqn{P_{NH}({\bar \nu_e}\to{\bar \nu_e})} \nonumber\\
&& =\,~~ 1 - 2 \, \sin^2\theta \cos^2\theta\,
\left( 1 - \cos \frac{ \Delta{m}^2_{\mbox{atm}} \, L }{ 2 \, E_{\nu} } \right)
\nonumber \\
&& - \,~~\frac{1}{2} \cos^4\theta\,\sin ^{2}2\theta_{\odot} \,
\left( 1 - \cos \frac{ \Delta{m}^2_{\odot} \, L }{ 2 \, E_{\nu} } \right)
\label{P21sol}  \\
& & +\,~~ 2\,\sin^2\theta\,\cos^2\theta\, \sin^{2}\theta_{\odot}\,
\left(\cos
\left( \frac
{\Delta{m}^2_{\mbox{atm}} \, L }{ 2 \, E_{\nu}} - 
\frac {\Delta{m}^2_{\odot} \, L }{ 2 \,
E_{\nu}}\right)
-\cos \frac {\Delta{m}^2_{\mbox{atm}} \, L }{ 2 \, E_{\nu}} \right)\, ,
\nonumber
\eea

%
\noindent where $E_{\nu}$ is the $\bar{\nu}_e$ energy.
If the neutrino mass spectrum is
with inverted hierarchy (IH), the $\bar{\nu}_e$
survival probability can be written in the form
\cite{BNPChooz, SPMPiai01,SChReactP}:

\bea
\lefteqn{P_{IH}({\bar \nu_e}\to{\bar \nu_e})} \nonumber\\
&& =\,~~ 1 - 2 \, \sin^2\theta\,\cos^2\theta\,
\left( 1 - \cos \frac{ \Delta{m}^2_{\mbox{atm}} \, L }{ 2 \, E_{\nu} } \right)
\nonumber \\
&& - \,~~\frac{1}{ 2} \cos^{4}\theta\,\sin ^{2}2\theta_{\odot} \,
\left( 1 - \cos \frac{ \Delta{m}^2_{\odot} \, L }{ 2 \, E_{\nu} } \right)
\label{P32sol}
  \\
& & + \,~~2\,\sin^2\theta\,\cos^2\theta\, \cos^{2}\theta_{\odot} \,
\left(\cos
\left( \frac
{\Delta{m}^2_{\mbox{atm}} \, L }{ 2 \, E_{\nu}} - 
\frac {\Delta{m}^2_{\odot} \, L }{ 2 \, E_{\nu}}\right)
-\cos \frac {\Delta{m}^2_{\mbox{atm}} \, L }{ 2 \, E_{\nu}} \right)\,. 
\nonumber
\eea

%
\noindent The $\bar{\nu}_e$
survival probability does not depend either 
on the angle $\theta_{atm}$
associated with the  atmospheric neutrino 
oscillations, nor on the CP violating phase 
$\delta$ in the PMNS matrix.

 In the convention (we will call A) in which 
the neutrino masses are
not ordered in magnitude and
the NH neutrino mass spectrum
corresponds to $m_1 < m_2 < m_3$, while 
the IH spectrum is associated with
the ordering $m_3 < m_1 < m_2$, 
it is natural to choose
\begin{equation}
\Delta m^2_{\odot}= \Delta m^2_{21} > 0~,~~~~~~~~~~~~~{\rm Convention~A}.
\label{A21sol}
\end{equation}
%
\noindent We can identify further
$\deltaatm$ with $\Delta m^2_{31}$
in the case of NH spectrum,
\begin{equation}
\deltaatm = \Delta m^2_{31} > 0~,~~~~~~~~~~~~~{\rm NH~spectrum~(A)}~,
\label{ANHdmatm}
\end{equation}
%
\noindent  and with
$\Delta m^2_{23} > 0$ if the spectrum is of the IH type,
\begin{equation}
\deltaatm = \Delta m^2_{23} > 0~,~~~~~~~~~~~~~{\rm IH~spectrum~(A)}~.
\label{AIHdmatm}
\end{equation}
%
\noindent In this convention 
the mixing angles in 
the standard parameterization
of the PMNS mixing matrix $U$ 
are given by: 
\begin{equation}
\theta_{12} = \theta_{\odot},~~~~~ 
\theta_{23} = \theta_{\mbox{atm}},~~~~~  
\theta_{13} = \theta~,~~~~~~~~~~~~~{\rm Convention~A}~. 
\label{Aangles}
\end{equation}
%

\begin{figure}[ht]
\begin{center}
\vspace{0.3cm} \epsfxsize = 8cm \epsffile{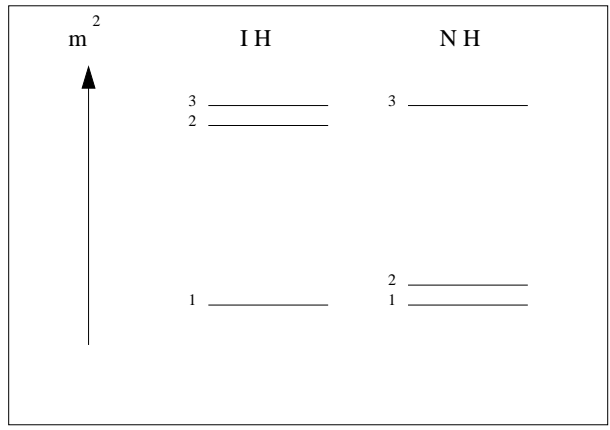}
\leavevmode
\end{center}
\caption{The two possible neutrino mass spectra, with Normal Hierarchy
(NH) and Inverted Hierarchy (IH), in the convention
in which $m_1 < m_2 < m_3$ (convention B). } 
\label{spectrum}
\end{figure}
%
\indent  One can also number (without loss of
generality) the neutrinos with definite mass 
in vacuum $\nu_j$ in such a way that their masses obey
$m_1 < m_2 < m_3$. In this alternative convention 
(we will denote as B)
it is convenient to choose
\begin{equation}
\Delta m^2_{\mbox{atm}} = \Delta m^2_{31} > 0~,~~~~~~~~~~~~~
{\rm Convention~B}~.
\label{Bdmatm}
\end{equation}
%
\noindent In the case of  NH 
neutrino mass spectrum we have 
\begin{equation}
\Delta m^2_{\odot}= \Delta m^2_{21} > 0~,~~~~~~~~~~~~~{\rm NH~spectrum~(B)}~,
\label{BNHdnsol}
\end{equation}
%
\noindent and 
\begin{equation}
\theta_{12} = \theta_{\odot},~~~~~ 
\theta_{23} = \theta_{\mbox{atm}},~~~~~  
\theta_{13} = \theta~,~~~~~~~~~~~~~{\rm NH~spectrum~(B)}~, 
\label{BNHangles}
\end{equation}
%
\noindent where $\theta_{ij}$ are the angles 
in the standard parameterization 
of $U$, \eq{Umix}. 
%

%
%
 If the neutrino mass spectrum is with IH,
one has in this convention 
(see, e.g., \cite{BGKP96,BPP1}):
\begin{equation}
\Delta m^2_{\odot}= \Delta m^2_{32} > 0~,~~~~~~~~~~~~~{\rm IH~spectrum~(B)}~,
\label{32sol}
\end{equation}

%
\noindent and

\begin{equation}
|U_{\mathrm{e} 2}| = \cos \theta_{\odot} \sqrt{1 - |U_{\mathrm{e} 1}|^2},
~~|U_{\mathrm{e} 3}| = \sin \theta_{\odot} \sqrt{1 - |U_{\mathrm{e} 1}|^2}~,
~~~~~~~~~~~{\rm IH~spectrum~(B)}.
\label{Ue23}
\end{equation}

%
\noindent
The mixing matrix element of $U$
constrained by the CHOOZ and Palo Verde data
is now $|U_{\mathrm{e} 1}|^2$ :
\begin{equation}
|U_{e1}|^2 = \sin^2\theta~,~~~~~~~~~~~~~{\rm IH~spectrum~(B)}~.
\label{Ue1th}
\end{equation}
%
\noindent We would like to emphasize 
that Eqs. (\ref{P21sol}) and 
(\ref{P32sol}) are valid in the two 
conventions for the ordering of 
the neutrino masses discussed above.

   Few comments are in order. The $\bar{\nu}_e$
survival probability
depends only on the four continuous parameters 
$\ms$, $\sss$, $\deltaatm$, $\sch$, 
and on a single ``discrete'' parameter --- the type
of the neutrino mass spectrum --- NH or  IH.
The terms in the second lines of 
\eq{P21sol} and \eq{P32sol} are responsible
for the dominant effects in the 
survival probability for the
intermediate baseline experiment 
under consideration in the case 
of the high-LMA solution
of the solar neutrino problem. 
The study of the dominant oscillations can thus constrain 
the solar neutrino oscillation parameters.
In a complete 3-neutrino oscillation 
analysis, one should also consider the
effect of the terms in the first and third lines of 
\eq{P21sol} and \eq{P32sol}. While the terms in the 
first lines give the 
sub-dominant, faster oscillations 
driven by $\deltaatm$, the terms in the last lines 
are responsible for the difference
between the probabilities  
corresponding to the NH and IH 
neutrino mass spectra. 
Both of them are suppressed by $\sch$. 
The detection of this sub-leading
oscillatory behavior, controlled by 
the atmospheric neutrino mass
squared difference, would be an 
indication of the non-vanishing of
the mixing angle $\theta$ in the PMNS matrix $U$. 
It will also enable the 
experiment under discussion 
to determine $\deltaatm$
with a high precision. Finally, for 
sufficiently large values of $\sch$ it should 
in principle be possible to distinguish 
the normal from the inverted hierarchy spectrum. 
Let us stress that the difference 
of the survival probabilities 
for the NH and the IH spectra lies in the
interference term: the existing solar neutrino 
and KamLAND data indicate
that $\theta_{\odot}$ is not maximal, so that 
$\sin \theta_{\odot} \neq\cos \theta_{\odot}$.
In ideal conditions in which
$\deltaatm$ is known to a very high precision,  
$\sch$ is not too small, $\sin^22\theta_{\odot}\ltap 0.90$,
and the statistics and the energy resolution of the
experimental apparatus are such to permit to
resolve the $\deltaatm-$driven oscillations, 
this could allow one to extract 
information about which
of the two hierarchical neutrino mass 
patterns is realized in nature.

One can rewrite the term in parenthesis
in the third line of \eq{P21sol} and \eq{P32sol} in the form:
\bea
\cos \left( \frac {\deltaatm \, L }{ 2 \, E_{\nu}} - \frac
{\Delta{m}^2_{\odot} \, L }{ 2 \,E_{\nu}} \right) -\cos \frac
{\deltaatm \, L }{ 2 \, E_{\nu}}\, &=& 2\,\sin  \, \frac
{\Delta{m}^2_{\odot} \, L }{ 4 \,E_{\nu}} \, \sin \,  \left( \frac
{\deltaatm \, L }{ 2 \, E_{\nu}} - \frac
{\Delta{m}^2_{\odot} \, L }{ 4 \,E_{\nu}} \right)
\label{interference}\eea
%
\noindent Thus, it consists of an oscillating function, with
approximately the same frequency as the 
$\deltaatm-$driven oscillations,
modulated with period half of the 
usual $\ms-$driven oscillations.
In particular, the amplitude of this 
term is the maximal possible
at values of $L/E_{\nu}$ where the 
survival probability goes to
its $\Delta{m}^2_{\odot}-$induced
minima, and vanishes at its local maxima.

\section{The experiment}
\label{section:numerical}

\begin{table}
\begin{ruledtabular}
\begin{tabular}{ccccc}
Power & Threshold & Baseline & Exposure & No osc \cr
P (GW) & $E_{th}$ (MeV)& L (km) & T (kTy) & Events\cr
\hline
5 &1.0& 20 & 3 & 14971 \cr
5 &2.6& 20 & 3 &  10585\cr
5 &1.0& 30 & 3 &  6654\cr
5 &2.6& 30 & 3 &  4704\cr
25 &1.0& 20 & 3 &  74855 \cr
25 &1.0& 30 & 3 &  33269\cr
\end{tabular}
\caption{\label{tab1}
Number of events in the case of absence of 
$\bar{\nu}_e$ oscillations for the different 
reactor experimental set-ups considered in this paper.}
\end{ruledtabular}
\end{table}
%
  Anti-neutrinos from nuclear reactor sources are detected
through their inverse $\beta$-decay reaction with protons in the
detector. The visible energy $E_{\mbox{vis}}$ of the
emitted positron is related to the energy of the 
incoming anti-neutrino $E_{\nu}$ and to the masses 
of the proton, neutron and positron as:
\bea
E_{\mbox{vis}}\,&\equiv&\, E_{\nu} \,+\, m_{e^+} \,-\,
(m_n-m_p)\,, \nonumber \\
          &\simeq& \, E_{\nu} \,-\,0.8\,\mbox{MeV}.
\eea
%
The $E_{\mbox{vis}}$  spectrum of 
detected events in the no
oscillation case, which takes 
into account the initial spectrum
of anti-neutrinos emitted by the 
reactor and the inverse $\beta$-decay 
cross section, has a bell-shaped distribution, 
centered at about $E_{\rm vis}\sim 2.8$ MeV. 
The total number of expected events for the no
oscillation case is given in Table \ref{tab1}.
We consider a single 
reactor plant with power of 
either 5 GW (achievable, e.g., 
at Heilbronn, Germany \cite{HLMA}) 
or 25 GW (achievable at Kashiwazaki, Japan),
as the only relevant source of 
neutrinos, neglecting the possible
contaminations due to other plants at 
larger distances. We also assume that the 
reactors have a 100\% efficiency.
It is trivial to adapt our results to 
a given reactor efficiency.
We present our results as function 
of the product of the exposure time
and the active mass of the detector. We will assume
for this product values in the range 
$3 \div 5 \,\mbox{kT}\,\mbox{y}$ 
and consider baselines $20 \div 30$ km.
The total statistics depends on 
${\cal L}=P~M~T$, the product of the reactor power ($P$),
the detector mass ($M$) and 
time of exposure ($T$), thus we express our
results in units of GWkTy.
Since the neutrino flux decreases as the 
inverse square of the baseline length, 
the shorter baselines obviously have 
much more statistics 
than the longer ones. We assume a 
liquid scintillation detector
similar to ones used in the 
other reactor experiments like
CHOOZ and KamLAND. We assume $8.48 \times 10^{31}$ free 
protons per kton of detector mass as in \kl \cite{KamLAND}.
We use an energy resolution of 
$\sigma(E)/E=5\%/\sqrt{E}$, $E$ in MeV 
for ``our'' detector, assuming some improvement with respect to
KamLAND, which reported a $\sigma(E)/E=7.5\%/\sqrt{E}$~\cite{KamLAND}.

  In Table \ref{tab1} we  
present the number of no oscillation events 
for two different visible energy 
thresholds of 1.0 MeV and 2.6 MeV. 
While at higher
energies the spectrum is known 
with relatively high accuracy, at lower
energies a possible contamination 
from geophysical neutrinos and
from the time variation of fuel composition is
expected \cite{Kopeikin:2001qv}. 
The KamLAND experiment
puts a conservative lower 
threshold of $E_{\mbox{vis}}\,>\,2.6$ MeV 
\cite{KamLAND} to avoid the error 
associated with the geophysical 
neutrinos. However, since the reactor 
flux is the highest around 2.8 MeV, 
it is desirable to include these 
parts of the energy spectrum 
to increase the statistical power 
of the experiment. We note from 
Table \ref{tab1} that the number of 
observed events goes up by 
a factor of 1.4 when the effective threshold 
energy $E_{th}$ is lowered from 
2.6 MeV to 1.0 MeV. Further, with a larger energy
window to be used in the data analysis, and
correspondingly a large interval in $L/E_{\nu}$,
it is possible, in principle, to reconstruct
a larger number of $\ma$-driven oscillation periods,
and even to detect both the SPMIN and SPMAX
connected to the $\ms$-driven oscillations.
The latter is crucial for the study of sub-leading
effects performed in the second part of this paper.  

    The ratio of the number of events 
expected from geophysics sources 
to the total number of antineutrinos 
detected can be estimated for 
a given experimental set-up.
In their first published data~\cite{KamLAND}, 
KamLAND estimated the 
number of these spurious events to be $9$, 
while the total data set 
contained 32 events with energy 
below the energy cut-off 
($E_{{th}} = 2.6$ MeV) and 54 above 
this threshold. For \kl this ratio therefore 
comes up to be about 30\% below 2.6 MeV. 
However, the number of background events
due to the geophysical neutrinos is 
quite model dependent and 
a very good understanding and modeling 
(see, e.g.,~\cite{Fiorentini:2003ww})
of the energy spectrum of these 
events would allow one to 
use the whole data sample in 
the range $(1.0 - 7.2)$ MeV, by 
subtracting the geophysics background from the 
observed event spectrum, with 
just an additional source of 
systematics in the analysis --
the extent of the uncertainty 
depending on the accuracy of the 
knowledge of the geophysical neutrino
background.

  The whole $e^+-$energy spectrum could 
also be used in a very 
high statistics experiment with 
a sufficiently powerful reactor source.
The contamination of the data sample from 
geophysical neutrino background 
at low energies is proportional 
to the active mass of the detector. 
Thus, decrease in the detector 
volume compensated with an increase 
in the reactor antineutrino flux, 
results in a reduction of the fraction 
of events due to this background. 
We can  determine the flux 
of $\bar{\nu}_e$ from a reactor
in the absence of oscillations 
as a function of the distance 
traveled and of the power of the source:
\bea
\Phi\,\equiv\,\frac{P}{L^2}\,.
\eea
%
For KamLAND, summing over all the reactors 
which contribute (see for instance~\cite{BGV} 
for a complete listing), one gets
\bea
\Phi^{\mbox{Kam}}\,\simeq\,3\,\times\,10^{-3}\,\mbox{GW}/{\rm km^2}\,.
\eea
%
As a comparison, using three 
possible choices for the pair 
$(L/\mbox{km},P/\mbox{GW})$ 
as $(30,5)$, $(20,5)$ and $(20,24)$, 
one gets for $\Phi$ values that
are bigger by a factor of $1.5$, $4$ 
and $20$, respectively.
Accordingly, the fraction of 
background events from the 
geophysical neutrinos 
can be reduced by 
this factor, as a consequence of
the larger flux. In particular, for
the last case, the geophysical neutrinos
contribution would be at the 
percent level even for a 1 kiloton 
detector like \kl 
and could be safely accounted for. 
For lower fluxes and large detectors, 
either a very accurate subtraction 
procedure needs to be 
applied, or it would be 
necessary to keep the 
cutoff in the low energy part of 
the spectrum as in KamLAND. 
In this paper we neglect throughout 
the contribution coming from 
geophysical neutrinos -- we either 
put an energy threshold of 2.6 MeV, 
or when we lower $E_{th}$ to 1.0 MeV 
we assume that either they can be 
accounted for, or they can be safely neglected.

   Backgrounds may also result 
from radioactive impurities and 
from cosmic ray interactions. However, 
these can be effectively 
suppressed in realistic experiments 
\cite{HLMA} and we completely 
neglect backgrounds in this paper.

  In presence of neutrino oscillation, 
the detected energy
spectrum at a distance $L$ from 
the reactor is obtained by 
convoluting the survival probability 
given by \eq{P21sol} or
\eq{P32sol} with the spectrum 
obtained with no oscillation. This
leads to a suppression of the number 
of detected events, and to a
distortion of the energy spectrum 
itself, whose dependence on the
parameters entering the survival 
probabilities may be used to
extract them through a $\chi^2$ fit of the data.

  The optimal distance in order to 
resolve the first minimum (SPMIN) of
the survival probability, and 
extract with good accuracy a
measurement of $\ms$ and $\sss$, 
according  to \eq{P21sol} and \eq{P32sol}, 
is given by
\footnote{This distance is also the optimal 
one for enhancing the effect
of the interference term in 
the survival probability
distinguishing the NH and IH 
spectrum cases, as seen from
\eq{interference}, because for 
this values of $L/E_{\nu}$ the
modulation of the beating-like term is maximal.} :
\bea
L^{\ast} \, \equiv \, \frac{2\pi E_{\nu}}{\ms }\,,
\label{Lstar}\eea
%
so that its choice depends on the 
actual value of $\ms$
and on the typical energy involved. 
If $E_{\nu}$ is at the
peak of the spectrum 
($3.5 \div 4$ MeV), 
this choice implies a
maximal depletion of the total 
number of events. We will consider 
always the case in which a relatively large
statistics is accumulated, so that 
the main source of information
is the differential energy spectrum 
of detected events. In this case
a large window of different energies 
can be used to determine
$L^{\ast}$. However, as long as the 
baseline of the experiment is 
large enough to include the first 
oscillation minimum (SPMIN) in the
interval of energies used for the
measurements, preferably in the 
part of the energy spectrum with 
highest statistics, the experiment 
is expected to display a good sensitivity 
to the solar neutrino oscillation parameters.

\begin{figure}
\begin{center}
\epsfxsize = 10cm
\epsffile{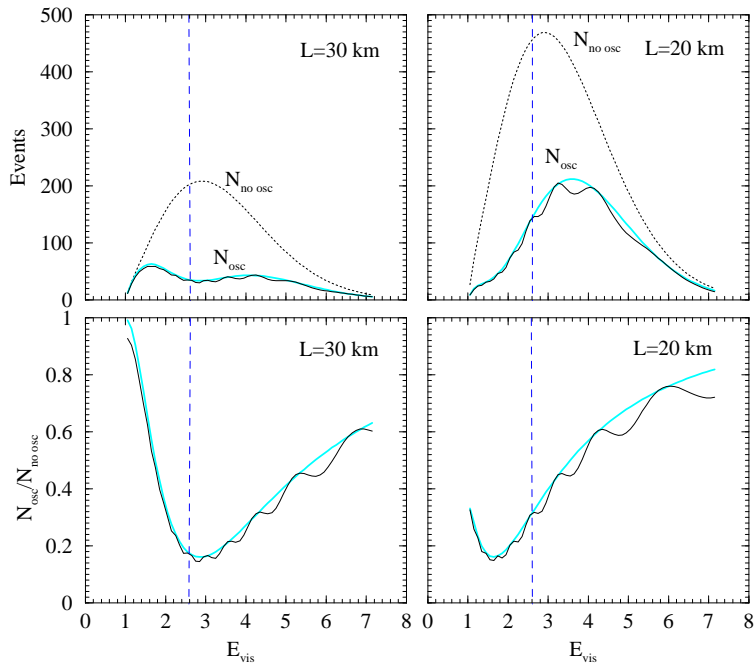}
\leavevmode
\end{center}
\caption{The observed $e^+-$energy 
spectrum (in 0.1 MeV bins) 
for the 20 km (right-panel) and 
30 km (left-panel) experiment 
with reactor power 5 GW and 
exposure of 3 kTy (15 GWkTy).
The dotted black lines in the 
top panels give the number of 
events in the absence of oscillations. 
The solid lines give the events 
for $\bar{\nu}_e-$oscillations with
$\ms=1.5\times 10^{-4}$ eV$^2$, 
$\sss=0.3$, $\deltaatm = 
2.5\times 10^{-3}$ 
eV$^2$. The solid cyan/grey line 
corresponds to the spectrum for $\sch=0.0$, 
while the solid black line shows the spectrum 
for $\sch=0.03$. The bottom panels 
show the corresponding ratio of events in
the cases of $\bar{\nu}_e-$oscillations 
and of absence of oscillations for 
the two values of  $\sch$.
The event spectra in the case 
of $\bar{\nu}_e-$oscillations
are for neutrino mass spectrum 
with normal hierarchy.}

\label{fig:spec}
\end{figure}

   We show in Fig. \ref{fig:spec} the 
spectrum of observed events 
for two specific baselines of 20 km 
and 30 km for reactor power 
of 5 GW and 3 kTy exposure. We note that 
the 30 km experiment has 
much less events than the 20 km experiment. 
The solid lines in the bottom panels 
show that the SPMIN for the 
30 km experiment comes around 2.8 MeV, 
where the flux is the highest.
For the 20 km experiment the SPMIN comes 
at around 1.6 MeV. 
Correspondingly, if the lower energy cut-off 
is set to $E_{th}=2.6$ MeV 
(shown by the dashed lines in Fig. 2), 
the 20 km experiment  
would miss the SPMIN. Therefore for the 
20 km experiment, the $E_{th}$ has 
to be lowered to below 1.6 MeV if 
the value of $\ms$ happens to 
be $1.5\times 10^{-4}$ eV$^2$. 
For higher values of $\ms$, the 
SPMIN will shift to higher energies 
and the threshold could be taken higher:
for  $\ms = 2.5\times 10^{-4}$ eV$^2$, for 
instance, SPMIN is at $E_{\rm vis} = 3.2~(5.3)$ MeV
for $L = 20~(30)$ km.
%

  Figure~\ref{fig:spec} also shows 
the ``sub-dominant'' oscillation effects 
dependent on $\deltaatm$ and $\sch$, 
imprinted on the large dominant 
oscillation wave driven by the 
solar neutrino oscillation parameters. 
The experiment could be expected to have 
sensitivity to the parameters 
$\deltaatm$ and $\sch$,
driving the sub-dominant oscillations, 
if the energy resolution 
of the detector would be good 
enough to enable a large statistics experiment 
like the one we consider here, 
to break the total number of events 
into fine energy bins. In particular, 
to resolve the oscillatory behavior 
one needs bins much smaller 
than the oscillation period 
\bea 
\Delta E\,\ll \, \Delta
E^{\ast}\,\equiv\,\frac{4\pi E^2_{\nu} } {\Delta m^2\,L}\,,
\label{resolution} \eea
%
\noindent where $\Delta m^2$ is the relevant 
neutrino mass squared difference 
driving the oscillations. 
This implies that for a given $L$,
higher energy resolution and smaller 
bins are required in order to resolve
effects due to a larger mass squared difference. 
As a numerical example, 
\bea 
\ms= 1.5 \, \times 10^{-4}\,\mbox{eV}^2 & \Longrightarrow &
\Delta E^{\ast}_{\odot} \,=\,7.1\,\mbox{MeV}\,,\nonumber
\\
\deltaatm = 2.7 \, \times 10^{-3}\,\mbox{eV}^2 
& \Longrightarrow & 
\Delta E^{\ast}_{\mbox{atm}} \,=\,0.4\,\mbox{MeV}\,,\nonumber
\eea
%
\noindent for energies at the peak of 
the spectrum ($E_{\nu} =
3.6$ MeV) and at the distance $L=30$ km. 
It is clear that energy bins with a 
width of 0.425 MeV (as in 
KamLAND) allow a good
reconstruction of the spectral 
distortion due to $\ms-$driven
oscillations, but would average 
out completely the sub-leading $\deltaatm$
effects. Using as a guide rule 
the assumption that to resolve an
oscillation one needs 
$\Delta E_{\mbox{vis}} \ltap \Delta E^{\ast}/4$, 
the reconstruction of 
$\deltaatm-$driven oscillations
requires to use energy bins   
having width of $\sim 0.1$ MeV . 
The use of shorter distances can 
soften this problem, as
\eq{resolution} indicates. However, 
with an energy resolution of 
$\sigma(E)/E=5\%/\sqrt{E}$, it should 
be possible to have bins of 
size of 0.1 MeV, at least as long as the 
energy is not very large. 
We will show that with bins of 0.1 MeV width, 
and $L=20$ km, one can improve on the CHOOZ 
limit for $\sch$, and if $\sch$ is 
found to be large enough to be detectable in this 
experiment, $\deltaatm$ can be determined with
a very high precision.

   With an experimental set-up 
which allows a sufficiently small bin size 
and large enough statistics to measure the 
sub-dominant oscillations, one can 
hope to gain information 
about the neutrino mass hierarchy 
using the last interference terms in 
\eq{P21sol} and \eq{P32sol}.

   Finally, one has to take into 
account the  presence of systematic 
errors both in the determination 
of the flux normalization 
and in the energy calibration of the 
detector. The KamLAND collaboration 
estimated a total systematic uncertainty of
about $6.42\%$ in their first published 
analysis~\cite{KamLAND}. The
bulk of this error comes from 
uncertainty in the overall 
flux normalization. The use of a near 
detector can improve this flux normalization 
error to as low as 0.8\% \cite{HLMA}. 
The error on energy calibration 
could be taken as 0.5\% \cite{Lindner}.
We use a total systematic uncertainty of $2\%$,
and we discuss in detail the effect of 
varying this uncertainty in the analysis which follows.



\section{Precision Neutrino Oscillation Physics : $\Delta m^2_{\odot}$ and 
$\sin^22\theta_{\odot}$ in the High-LMA Region}
 \label{section:solar} 

  In this Section we study quantitatively 
the precision measurements 
of the solar neutrino oscillation parameters
$\ms$ and $\sss$,
possible in the intermediate baseline 
reactor experiment discussed 
in the previous section under the 
assumption that $\ms$ lies in the high-LMA region. 
We simulate the ``data'' at certain 
plausible values of $\ms$ and $\sss$ and 
obtain allowed regions in the parameter 
space from a dedicated $\chi^2$ analysis. 
For the errors we assume a
Gaussian distribution and define our $\chi^2$ as 
\bea
\chi^2 =  \sum_{i,j}(N_{i}^{data} - N_{i}^{theory})
(\sigma_{ij}^2)^{-1}(N_{j}^{data} - N_{j}^{theory})~,
\label{chig}
\eea
%
\noindent where $N_i^\alpha$ $(\alpha=data,theory)$ 
is the number of events in 
the $i^{\rm th}$ bin, $ \sigma_{ij}^2$ 
is the covariant error matrix 
containing the statistical and systematic 
errors and the sum is over all 
bins. We compare the sensitivities obtained 
for two possible baselines, 20 km and 30 km. 
To quantify the sensitivity we define the relative precision
$p_a$ for a certain parameter $a$ at a given C.L. as
\bea
 p_a = \frac{ a_{max} - a_{min}}
{a_{max} + a_{min}}~,
\label{error}
\eea
%
\noindent where
$a_{max}~(a_{min})$ are the maximal (minimal) allowed
value of $a$ found at the chosen C.L.
We check quantitatively
the impact of (1) the energy 
threshold $E_{th}$, (2) bin size $\Delta E$ 
and (3) systematic uncertainties. 
We also show the impact of increasing 
the statistics by a factor of 5.

   As has been noted earlier in the context of \kl \cite{cc},
the sensitivity to $\sss$ can be reduced considerably
by the uncertainty in the parameter $\sch$.
We therefore analyze first the impact this 
uncertainty can have on the precision of $\sss$ 
determination in the
intermediate baseline reactor experiment
of interest.
Since the baseline is optimized 
to measure predominantly the 
$\ms-$driven oscillations, we assume that 
the $\ma-$driven oscillations 
are averaged out.
This corresponds to the realistic case of
using a sufficiently large $e^+-$energy bin size. 
The expression for the survival probability 
would then reduce to
\bea
P_{ee}
\approx\cos^4\theta
\left( 1 - \sin ^{2}2\theta_{\odot}
\sin^2\frac{ \Delta{m}^2_{\odot} \, L }{ 4 \, E_{\nu} }\right)~,
\label{pee}
\eea
%
\noindent where we have neglected the term $\sim \sin^4\theta$.
Therefore the uncertainty in $\sch$ essentially 
brings up to a $\sim 10\%$ 
uncertainty in the $\bar{\nu}_e$ survival probability.
Since the factor $\cos^4\theta$ can only 
reduce the survival probability, it does not affect the upper limit 
on the allowed range of $\sss$. However,
it can have an effect on the lower limit on $\sss$
reducing it further, and thus 
can worsen, in principle, the precision of the experiment.
Using eq. (\ref{pee}) we get approximately for
%
%
%
\noindent 
the {\it additional} 
error on $\sin^22\theta_\odot$ due to
the uncertainty in the value of $\sch$ 
\bea
\delta(\sin^22\theta_\odot)\approx 
\frac{2\Delta P_{ee}\sin^2\theta}
{\sin^2\frac{ \Delta{m}^2_{\odot} \, L }{ 4 \, E_{\nu} }}
+ 2~\frac{(1 - \sin^2 2\theta_\odot 
~\sin^2\frac{ \Delta{m}^2_{\odot} \, L }{ 4 \, E_{\nu} })
~\Delta (\sin^2\theta)}
{\sin^2\frac{ \Delta{m}^2_{\odot} \, L }{ 4 \, E_{\nu} }}~,
\label{extra}
\eea
%
\noindent
where $\Delta P_{ee}$ and $\Delta (\sin^2\theta)$ are the 
uncertainties in the determination of the survival probability 
and $\sch$, respectively.
We work in a scenario in which one would use the 
SPMIN in order to determine $\sin^22\theta_\odot$.  
In the SPMIN region one has
$\sin^2(\Delta{m}^2_{\odot}L/ 4E_{\nu})\sim 1$ 
and therefore
\bea
\delta(\sin^22\theta_\odot)\approx 
2\Delta P_{ee}\sin^2\theta
+ 
2~\cos^2 2\theta_\odot ~\Delta (\sin^2\theta)~.
\label{extraour}
\eea
%
Thus, in this SPMIN scenario,
the first term gives an extra contribution of 
about $2\Delta P_{ee}\sin^2\theta$
{\it to the allowed range of $\sin^22\theta_\odot$}. 
Since, as we will see in the present Section, 
in the experiment under consideration the
allowed range of $\Delta P_{ee} \ltap 0.1$ even under the 
most conservative conditions, this term gives an extra contribution 
to the allowed range which is $\ltap 0.01$. The second term is 
independent of the precision of a given experiment 
(i.e., on $\Delta P_{ee}$)
and depends only on the best-fit 
value of $\cos^2 2\theta_\odot$ and on 
the error in $\sch$.
For the current $3\sigma$ error 
in $\sch$ of 0.05  and 
best-fit $\cos^2 2\theta_\odot$ of $0.16$, 
this would give an increase of 
only $0.018$. The suppression of 
this term is mainly due to the 
presence of the factor $\cos^2 2\theta_\odot$
 which is a relatively small number
for the current best-fit solution. 
Thus, even though the uncertainty in $\sch$ brings a
10\% uncertainty in the 
value of $P_{ee}$, it increases the allowed range 
of $\sin^22\theta_\odot$ only by $\ltap (2-3)\%$,
if one uses the SPMIN region for the
$\sin^22\theta_\odot$ determination.

  The error in the measured value of 
$\sin^2 2\theta_\odot$ due to the uncertainty in   
the value of $\sch$ is considerably larger 
in an experiment like \kl, which is sensitive 
primarily to the region of the 
maximum of the $\bar{\nu}_e$ survival probability
\footnote{This is valid both for
$\ms \sim 7.2 \times 10^{-5}$ eV$^2$ and for
$\ms \sim 1.5 \times 10^{-4}$ eV$^2$.}
(SPMAX). In this case the oscillatory term
$\sin^2(\Delta{m}^2_{\odot}L/ 4E_{\nu})\neq 1$ and 
can be quite small, so that the extra 
contribution to the uncertainty in 
$\sin^22\theta_\odot$, given roughly by \eq{extra}, 
would be rather large. The first term in the 
expression in the right-hand side of \eq{extra}
becomes relatively large due to presence 
of the oscillatory term in the 
denominator and the second term becomes even larger due 
to the presence of the same term 
both in the denominator and in the numerator.
Thus, for \kl the impact of the uncertainty in 
$\sch$ on $\sss$ determination 
is essentially related to the fact 
that the $\ms-$dependent oscillatory 
term for \kl is relatively small, 
resulting in rather big contributions 
from both terms in \eq{extra}.
This explains the relatively large effect 
of the $\sch$ uncertainty on the
$\sin^22\theta_\odot$ determination
noticed in \cite{cc}. 
We assert that to reduce the impact of the $\sch$ 
uncertainty on the determination of $\sss$, 
one needs to ``tune'' the experiment 
to the SPMIN. This observation gives further 
credence 
to our statement 
that the best experimental 
set-up for determining the solar 
neutrino mixing angle with high precision 
is a reactor experiment sensitive to 
the SPMIN.  

\begin{figure}
\begin{center}
\epsfxsize = 11cm
\epsffile{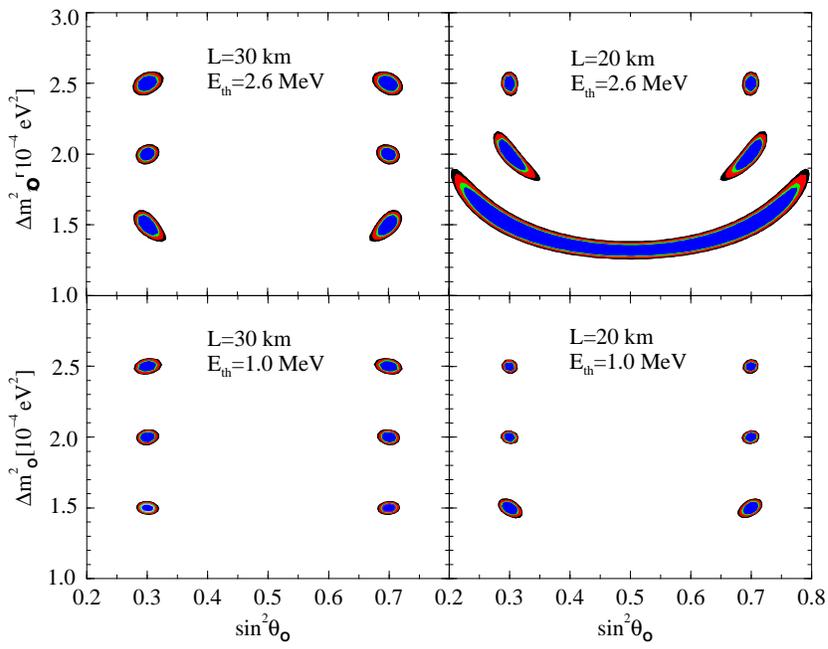}
\leavevmode
\end{center}
\caption{Simulated two parameter fit 
of the $e^+-$energy spectrum in the 
$\ms -\sss$ plane. 
We assume $\sch=0$, $\sss=0.3$ and vary 
the value of $\ms$. 
We compare results at $L=20$ km 
and  $L=30$ km, with 
the lower cut-off on 
the energy spectrum at
$E_{{th}}=2.6$ MeV
and at $E_{{th}}=1.0$ MeV.
The energy bin size is chosen as $\Delta E = 0.425$ 
MeV, systematic uncertainties 
are taken as $2\%$ and the statistics correspond to
$15$ GWkTy. Shown are the 
$90\%$, $95\%$, $99\%$ and $99.73\%$ 
($3 \sigma$) C.L. contours.}
\label{fig:solLEth}
\end{figure}
\begin{figure}
\begin{center}
\epsfxsize = 11cm
\epsffile{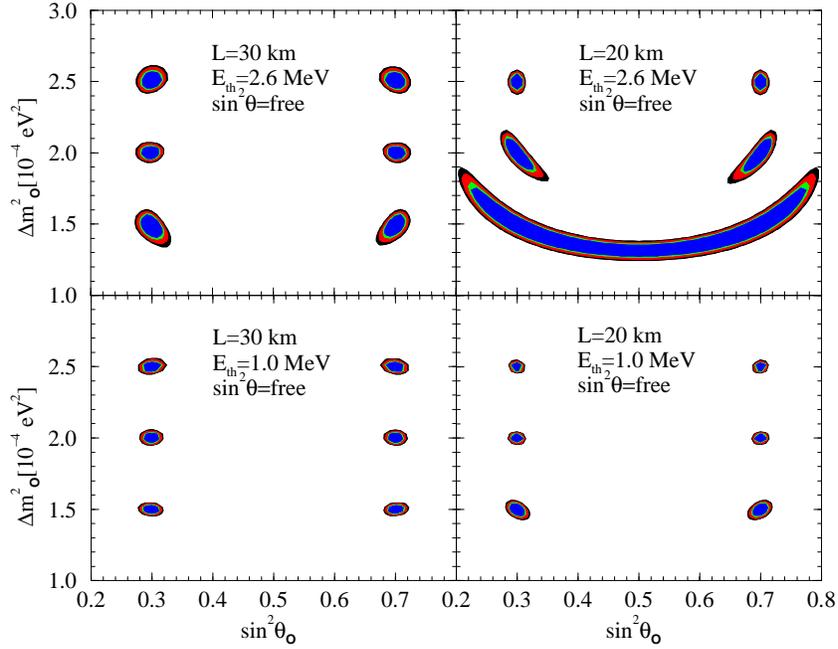}
\leavevmode
\end{center}
\caption{
The same as in Fig.~\ref{fig:solLEth} but for $\sch$ 
allowed to vary freely
within its 99.73\% C.L. allowed range, $\sch < 0.05$.}
\label{fig:solLEththfree}
\end{figure}
\begin{figure}
\begin{center}
\epsfxsize = 10cm
\epsffile{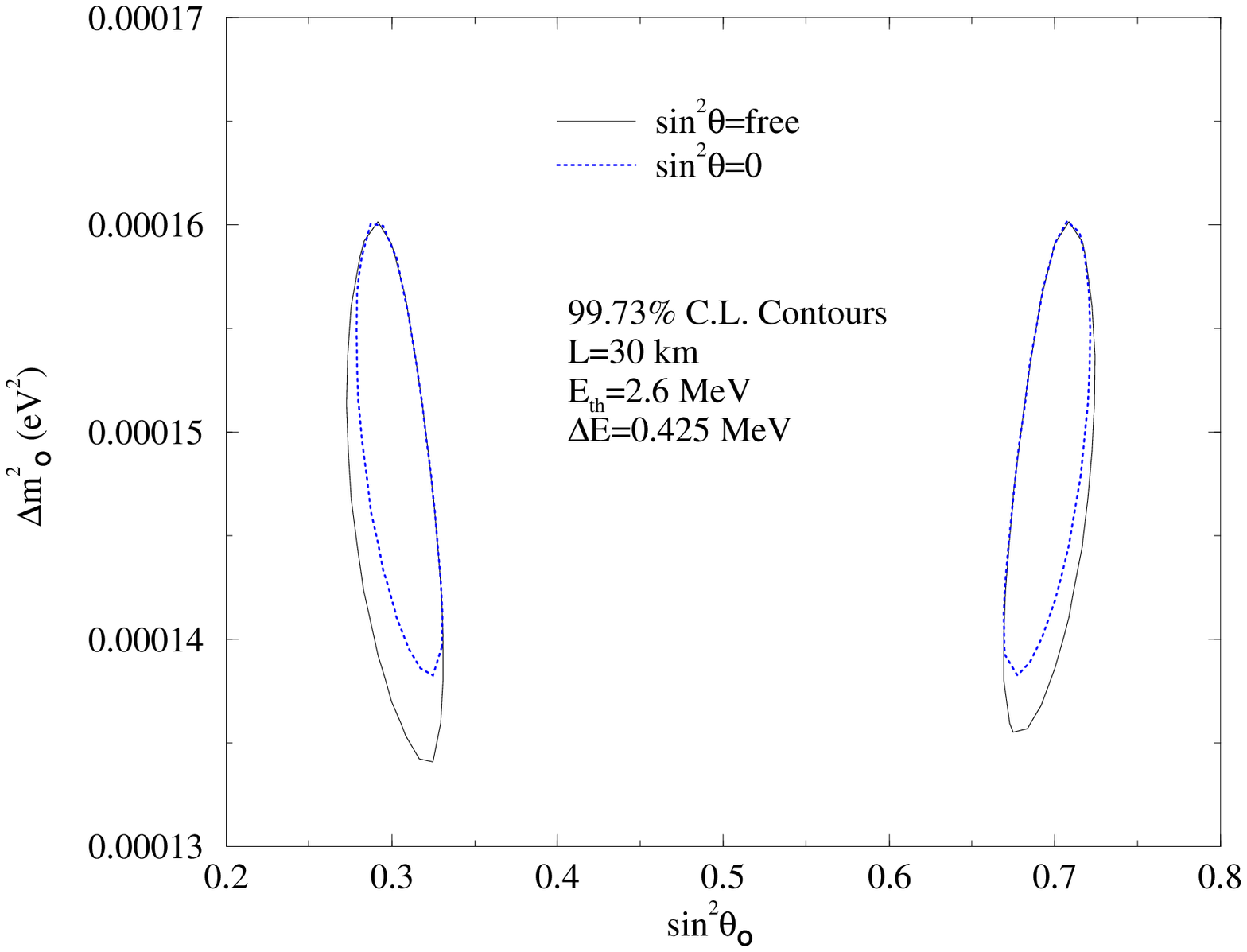}
\leavevmode
\end{center}
\caption{
Comparison of the allowed regions obtained at the 99.73\% C.L.
for $\sch = 0$ and allowing $\sch$ to vary
freely within its 99.73\% C.L. allowed range,
$\sch < 0.05$. The curves correspond to 
the top left-hand panels of Figs.~\ref{fig:solLEth} and 
\ref{fig:solLEththfree} and ``true'' $\ms=1.5\times 10^{-4}$ eV$^2$.}
\label{fig:blow}
\end{figure}

  In Figs.~\ref{fig:solLEth} and 
\ref{fig:solLEththfree} we show 
the regions of allowed values of  
$\ms$ and $\sss$,  obtained with 
``data'' simulated at  various values of 
$\ms > 1.0 \times 10^{-4}$ eV$^2$ and $\sss=0.3$.
We assume a statistics corresponding to an 
exposure of 15 GWkTy (5GW$\times$3kTy) 
\cite{HLMA}, bin size 
$\Delta E=0.425$ MeV and plausible 
systematic uncertainty of 2\%.
Figure~\ref{fig:solLEth} shows the allowed regions
obtained for $\sch = 0$, corresponding 
to the case of 2-neutrino  $\ms-$ driven oscillations. 
In Fig.~\ref{fig:solLEththfree} we display
the corresponding allowed regions in a complete 
3-neutrino mixing scheme 
with $\sch$ allowed to vary freely
within the 99.73\% C.L. allowed range,
$\sch < 0.05$. As a comparison 
of the two figures indicates,
the effect of keeping $\sch$ free, 
on the allowed ranges of the solar neutrino 
oscillation parameters 
in general, and on $\sss$ in particular, 
is rather small. This is 
compatible with the conclusions of the 
analysis presented above. 
To further substantiate our point 
pertaining to the impact of the $\sch$ uncertainty, we show 
in Fig.~\ref{fig:blow} a blow-up 
of the 99.73\% C.L. contours for the 
$\sch$ free (within the bound $\sch < 0.05$) and the $\sch=0$ cases, 
obtained with $L=30$ km, $E_{th}=2.6$ MeV and with 
assumed ``true'' value of $\ms=1.5\times 10^{-4}$ eV$^2$.
We stress that the effect of keeping 
$\sch$ free is small and henceforth in the 
rest of this Section neglect the effect of 
$\sch$ uncertainty on the precision of
the $\ms$ and $\sss$ determination.

  In Fig.~\ref{fig:solLEth} we compare 
the precision obtained 
on the solar neutrino oscillation 
parameters for two 
intermediate baselines of 20 km and 30 km, 
with and without imposing 
the low energy cut-off of 2.6 MeV 
\footnote{For all the cases we get 
two solutions in $\sss$, the real one 
and a ``fake'' one on the ``dark-side'' 
($\sss > 0.5$). This fake ``dark-side'' solutions  
are ruled out by the solar neutrino data. 
Nonetheless we show them 
in our plots for the sake of completeness.}. 
For the baseline of $L=30$ km, 
depending on the ``true'' values of the 
parameters, at 99\% C.L. a 
precision respectively of 
$(6 - 7)\%$ and $(3 - 6)\%$ is possible 
to achieve for $\sss$ and $\ms$ even 
with $E_{th}=2.6$ MeV. These
errors reduce only slightly 
to $(5 - 6)\%$ and $\sim 2\%$,
respectively, 
when the threshold is lowered to 1.0 MeV, 
owing mainly to the fact that the statistics 
increases, but also to the fact 
that the SPMIN is fully used in the measurement.

   For the $L=20$ km experiment 
on the other hand, the precision for both 
$\ms$ and $\sss$ (and especially for $\sss$) 
is relatively low if the effective threshold of 
$E_{th}=2.6$ MeV is applied, and if 
$\ms = 1.5 \times 10^{-4}$ eV$^2$. As discussed
in the previous Section, 
for larger values of $\ms$ which produce the SPMIN 
at $E_{\rm vis} > 2.6$ MeV,
it is still possible to 
reach high precision in the determination
of $\sss$ and $\ms$
even with the cut-off
of $E_{th}=2.6$ MeV. These values 
are somewhat disfavored by the current 
data~\cite{SolFit1,SolFit2,SolFit3}.
If the threshold is lowered to 1.0 MeV, 
the experiment can ``see'' the full 
SPMIN even for $\ms = 1.5 \times 10^{-4}$ eV$^2$
and the sensitivity to 
both $\ms$ and $\sss$ becomes remarkable. 
The $\ms$ can be determined with a
$\sim 2\%$ precision, while the $\sss$ 
would be known to within  $\sim (3-4)\%$,
at the 99\% C.L.
With $E_{th}=1.0$ MeV even though both 
$L=20$ km and $L=30$ km baseline 
experiments would be sensitive to SPMIN, 
the precision of the shorter 
baseline experiment is higher, owing 
to the larger statistics expected. 

\begin{figure}
\begin{center}
\epsfxsize = 11cm
\epsffile{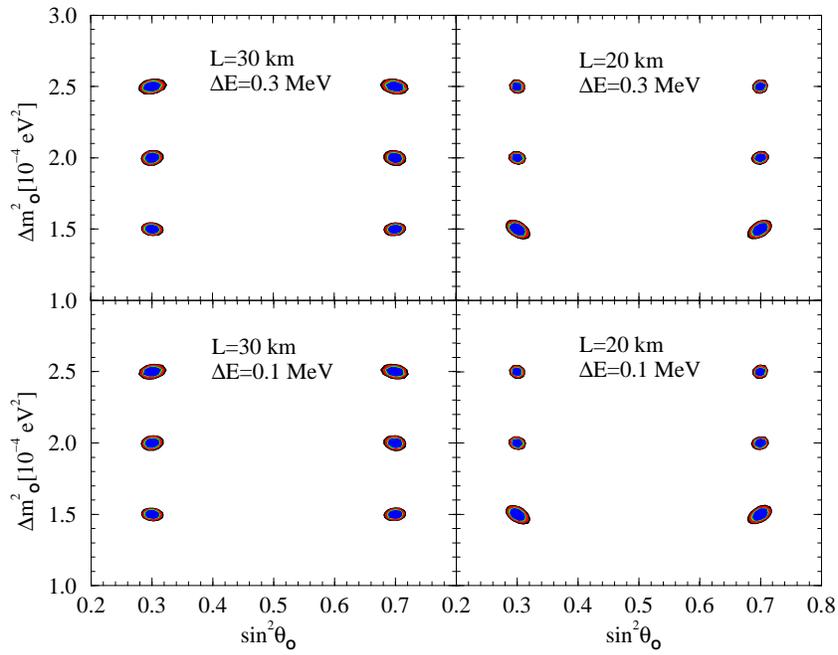}
\leavevmode
\end{center}
\caption{The same as the lower panels in
Fig.~\ref{fig:solLEth}, but showing the 
effect of reducing the $e^+-$energy bin size $\Delta E$.
}
\label{fig:solBin}
\end{figure}
%

In Fig.~\ref{fig:solBin}, we study 
the effect of using smaller bins
in the $e^+-$energy spectrum. As explained 
in the previous Section, there
is no substantial improvement in 
the precision of the fit
to the solar neutrino oscillation  
parameters $\ms$ and $\sss$ with 
this strategy. The binning energy 
used by KamLAND ($\Delta E=0.425$ MeV) 
is already sufficient and one does 
not need to use smaller bins.
However, we will see in the next sections that 
a better energy resolution could
allow for the extraction of other 
information embedded in the sub-dominant 
oscillations.
\begin{figure}
\begin{center}
\epsfxsize = 11cm
\epsffile{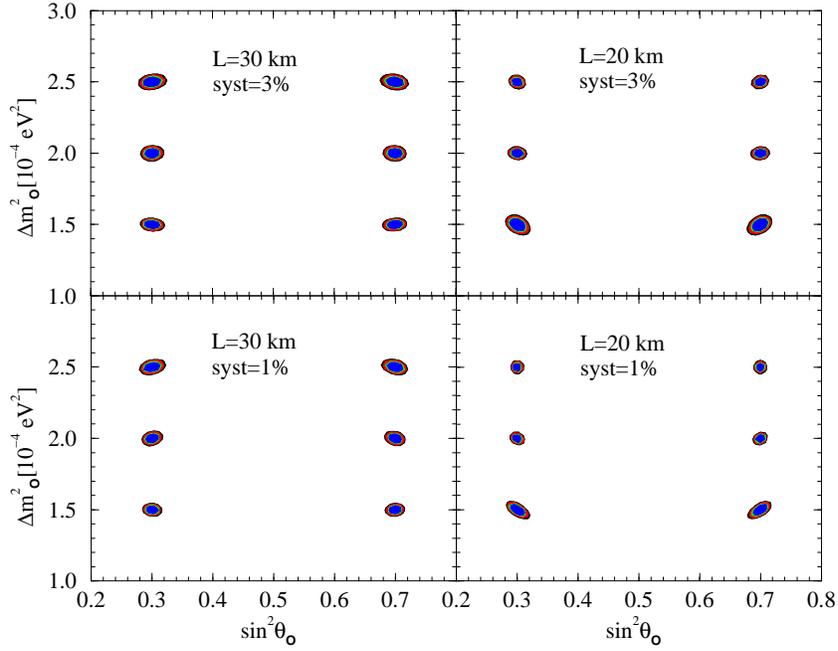}
\leavevmode
\end{center}
\caption{The same as the lower panels in 
Fig.~\ref{fig:solLEth} but showing 
the effect of varying the systematic 
error.
}
\label{fig:solSys}
\end{figure}
%

   One of the crucial issues in these 
types of experiments is thought to be  
the control of the systematic error. Indeed, 
a large systematic uncertainty on the 
energy spectrum could threaten to wash out the 
possibility of extracting any information from the data.
In Fig.~\ref{fig:solSys} we show the 
impact of the systematic 
uncertainty on the expected precision.
The plots show the allowed areas 
obtained assuming 1\% and 3\% 
systematic uncertainties and can be 
compared with the lower panels 
in Fig.~\ref{fig:solLEth}, where the 
systematic uncertainty is taken 
as 2\%. By comparing the relative 
precision on the parameters, we 
note that while the precision
on $\ms$  remains essentially unaffected 
by the exact value of the systematic uncertainty
if the latter does not exceed $\sim4\%$, 
the precision of $\sss$ may show a 
very mild dependence, the error in 
$\sss$ obviously decreasing with 
the reduction in the systematic error. 
However, the impact is not large 
(see also \cite{Ch} where the 
performance of \kl is studied under 
assumption of various anticipated 
systematic errors). 
This implies that, at least for 
reasonably small values of the 
systematic uncertainties, their 
impact on the precision of 
the parameter determination 
is only marginal.
This result is essentially due to 
the fact that we are assuming rather 
large statistics, allowing for a 
good reconstruction of the 
oscillatory pattern in the energy 
spectrum. Similar conclusions 
were recently drawn in the context 
of a short baseline experiment at 
nuclear reactors ($L=1.7$ km) in~\cite{Lindner}.

\begin{figure}
\begin{center}
\epsfxsize = 11cm
\epsffile{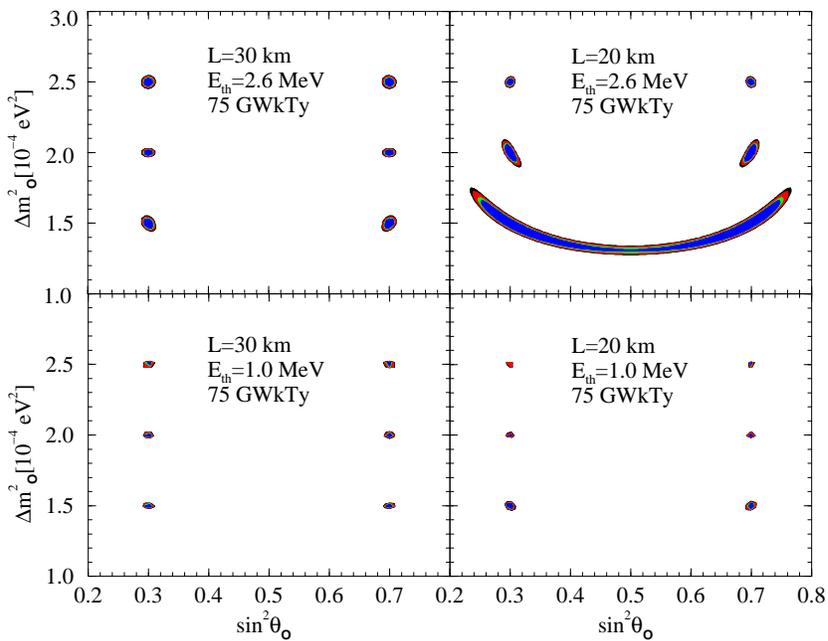}
\leavevmode
\end{center}
\caption{The same as  
in Fig.~\ref{fig:solLEth}, but showing 
the effect of increasing the statistics 
five times, corresponding to 75 GWkTy.
}
\label{fig:solStat}
\end{figure}
%

  By contrast, we illustrate 
in Fig.~\ref{fig:solStat} 
the huge effect that 
an increase in the statistical sample  
could lead to. We show 
the allowed regions, corresponding 
to Fig.~\ref{fig:solLEth}, 
that would be obtained
if the statistics were increased 5 fold. 
This could be done using 
a reactor complex of total power 
comparable to the Kashiwazaki 
site ($24.3$ GW) and a large enough detector.
We note that the error in 
the statistical determination of both 
$\ms$ and $\sss$ would go down 
below the percent level. 
With this statistics even the effect 
of the systematic uncertainty on 
the precision determination 
of the parameters 
is expected to be low and one can 
achieve remarkable accuracy.

\section{Precision Physics from 
$\deltaatm-$Driven (Sub-Dominant) Oscillations}
\label{section:subdom}

   In this Section we will consider 
the set-up where the experiment 
could be sensitive to the $\deltaatm-$
driven 
sub-dominant oscillations.
As discussed in Section~\ref{section:numerical}, 
one  has to use relatively small energy 
bins in order to observe 
these oscillation effects.
We first consider the 
potential of the intermediate baseline 
reactor experiment of interest
in improving the bound on $\sch$. We next assume 
that $\sch$ is sufficiently large and can 
therefore lead to observable 
$\deltaatm-$driven oscillations and
investigate under what conditions one can 
use these ``fast'' oscillations 
to determine $\deltaatm$ with a high precision. 
Finally, we study the possibility of achieving 
what is probably a most ambitious goal 
-- to get information on the neutrino 
mass hierarchy by observing the
reactor $\bar{\nu}_e$ 
oscillations at intermediate baselines.

\subsection{Improving the Limit on $\sch$}
\label{section:lma1}

\begin{figure}
\begin{center}
\epsfxsize = 11cm 
\epsffile{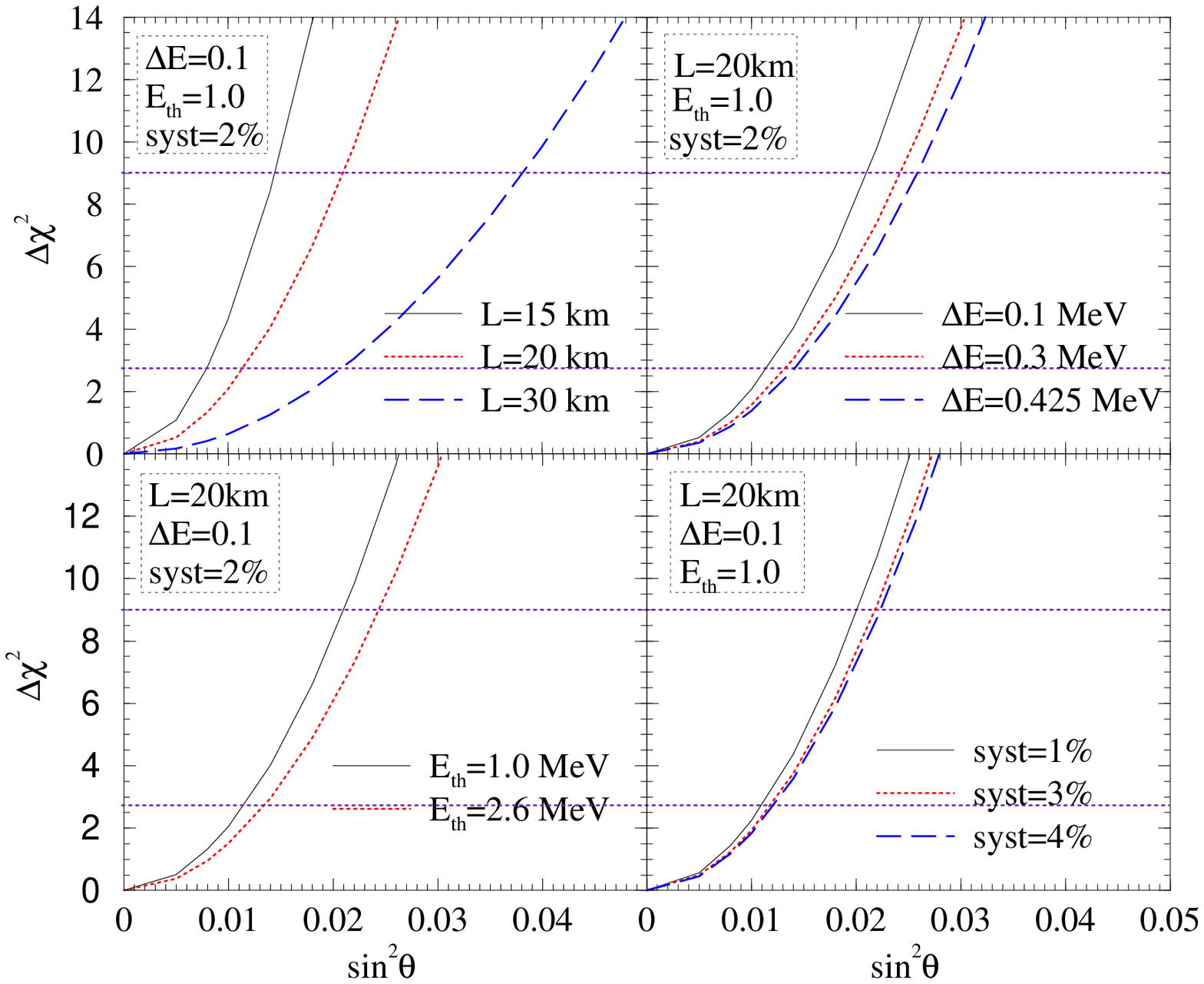}
\leavevmode
\end{center}
\caption{Study of the sensitivity to $\sch$. 
Shown are the $\Delta \chi^2$ obtained 
as a function of $\sch$,
when fitting a simulated ``data'' 
set generated for $\sch=0$,
$\ma=2.5\times 10^{-3}$ eV$^2$, 
$\ma = 1.5\times 10^{-4}$ eV$^2$ and 
$\sss = 0.3$. For a particular value of 
$\sch$, the other parameters are allowed 
to vary freely in the fit.
The two horizontal lines correspond 
to $3 \sigma$ and $90 \%$ C.L., respectively.
We assume a total exposure of 
${\cal L}=15$ GWkTy, and explore the impact
of the baseline ($L$), width of  energy bins ($\Delta E$), 
the systematic uncertainties and the 
application of a low-energy cut-off ($E_{{th}}$) on the 
$\sch$ sensitivity.
}
 \label{fig:theta}
\end{figure}

 We investigate in what follows the
impact of several experimental 
conditions on the achievable sensitivity on $\sch$.
We show in Fig.~\ref{fig:theta} the $\sch$ 
sensitivity obtainable 
in a 15 GWkTy statistics experiment.  
We generate the ``data'' at $\sch=0$ and at 
each value of $\sch$ we 
plot the difference between the 
$\chi^2$ function obtained at
that value of $\sch$ and the $\chi^2_{min}$, 
which obviously comes at 
$\sch=0$. We define the $3\sigma$ and 
90\% C.L. limit levels as 
$\Delta \chi^2=9$ and 2.71, respectively, 
corresponding to a one parameter fit. 

 The sensitivity to $\sch$ increases substantially
if the baseline $L$ is reduced.
It follows from \eq{Lstar} that 
the best sensitivity to $\sch$ 
is achieved for $L=$ few km \cite{mina,Lindner}. 
The intermediate baselines we are considering 
are not optimized for the measurement of $\theta$.
The sensitivity to $\sch$ of the experiment 
under discussion is 
worse than that of the experiment proposed 
recently in~\cite{mina,Lindner} 
with a distance of $1.7$ km. Still, 
the experiment with $L=20$ km, would allow 
to put an upper bound of 
\bea
\sch\,<\,0.021~(0.012)\,,
\eea
%
\noindent at the $3 \sigma~(90\%)$ C.L.
(Fig.~\ref{fig:theta}, top left panel),
which is a noticeable improvement over the 
current $3\sigma$ bound of $\sch<0.05$.

   Rather relevant for the sensitivity 
to $\sch$, in addition to the distance $L$, 
is the width of the final state $e^+$ energy bins. 
As could be expected, the smaller bins 
give a better sensitivity to the value of $\sch$.
Bins with width of 0.1 MeV 
give a 50\% improvement 
in sensitivity compared to 
the case of bin width of 0.425 MeV.
Another relevant factor is the 
presence of a low energy cut-off in the 
part of the spectrum used in the fit:
the sensitivity improves with the inclusion 
of the lower energy data by, e.g., changing the cut-off
energy from 2.6 MeV to 1.0 MeV.
Finally, the last panel in Fig.~\ref{fig:theta}  
illustrates the effect of 
the systematic uncertainties 
on the sensitivity to $\sch$. 
As discussed in detail in the previous Section,  
the systematic error in the 
$\sch$ measurement has a small 
but non-negligible effect -- 
the limit improving 
marginally with the reduction of the error.

\begin{figure}
\begin{center}
\epsfxsize = 11cm 
\epsffile{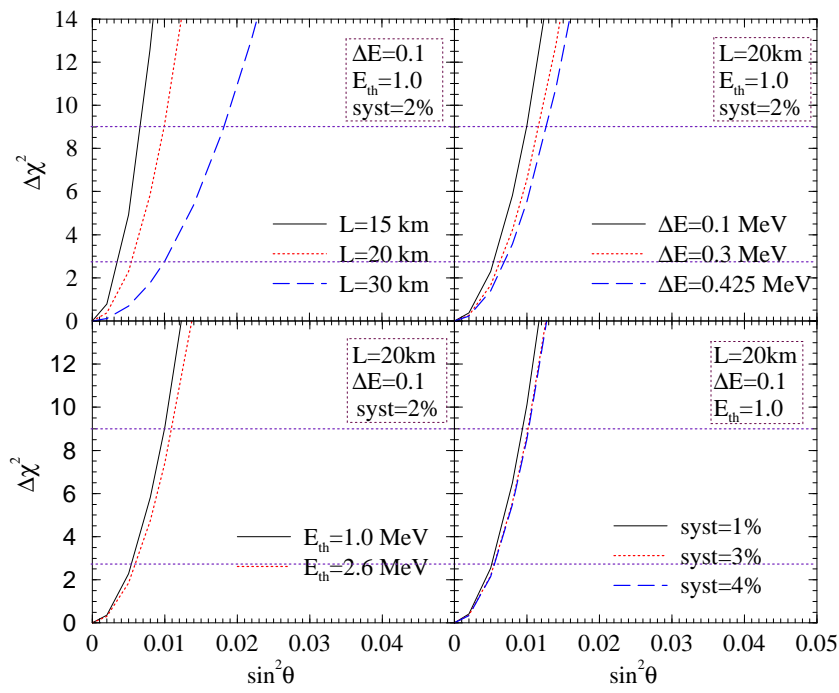}
\leavevmode
\end{center}
\caption{The same as in Fig.~\ref{fig:theta}, 
but with statistics increased 
by a factor of five (${\cal L}=75$ GWkTy).}
 \label{fig:theta25}
\end{figure}

    We show in Fig.~\ref{fig:theta25} the $\sch$
sensitivity when the statistics 
are increased by a factor of five:
we consider statistics corresponding to 
75 GWkTy. Clearly, the increase in statistics 
improves the $\sch$ sensitivity 
and at $3\sigma~(90\%~{\rm C.L.})$ limit 
for the $L=20$ baseline case reads (Fig.~\ref{fig:theta25}, 
top left panel)
\bea
\sch\,<\,0.01~(0.0055)\,.
\eea
%
\noindent Thus, for these very high statistics 
we get a  sensitivity to $\sch$ which is 
almost of the same order as that expected 
to be reached  
in the ``Reactor-I'' experiment 
discussed in \cite{Lindner}.
Note that with the increase of the
statistics, the effect of both the systematic 
error and of the value of $E_{th}$ chosen 
on the $\sch$ sensitivity 
decreases. The decreasing of the 
energy bin size, however, 
has essentially the same effect 
of increasing the sensitivity.


\subsection{Measuring 
\deltaatm~~and $\sch$}
\label{section:atm}



\begin{figure}
\begin{center}
\epsfxsize = 11cm 
\epsffile{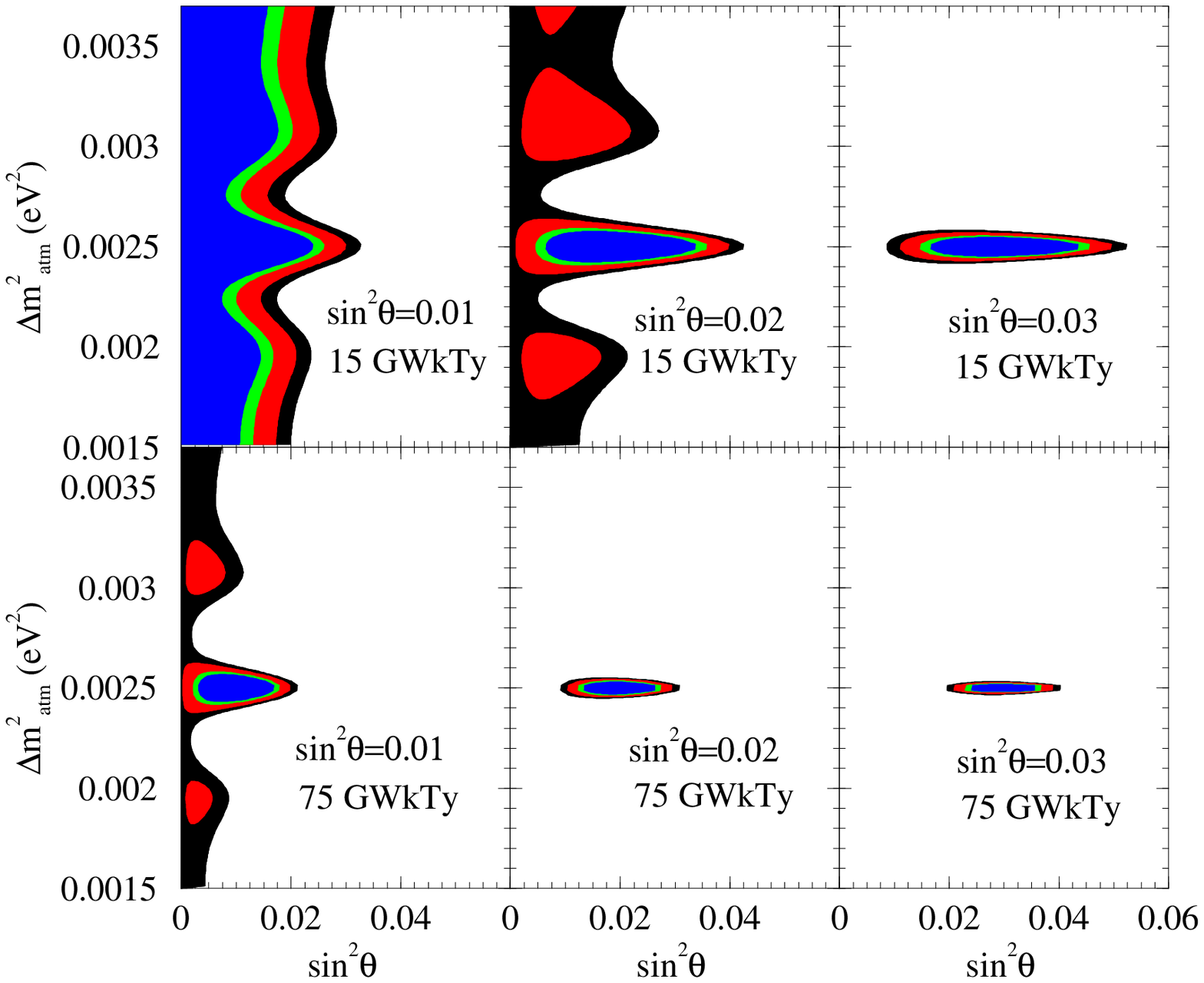}
\leavevmode
\end{center}
\caption{The 90\%, 95\%, 99\% and 99.73\% C.L. 
allowed regions obtained 
by fitting ``data'' set simulated 
at $\ma=2.5\times 10^{-3}$ eV$^2$ and 
three different ``true'' values 
of $\sch$. The solar parameters 
$\ms$ and $\sss$ are taken to 
be $1.5\times 10^{-4}$ eV$^2$ and 
$0.3$, respectively. The statistics 
assumed corresponds to ${\cal L}=15$ GWkTy for 
the upper panels and to ${\cal L}=75$ GWkTy 
for the lower panels. The distance used is $L=20$ km.
The bin size is taken as 0.1 MeV, the lower 
energy cut-off is $E_{th}=1.0$ MeV, 
and the systematic uncertainty is 
assumed to be 2\%. Both the 
``data'' set and the fitted 
$e^+-$spectrum correspond to NH 
neutrino mass spectrum.
}
\label{fig:d31theta}
\end{figure}
\begin{figure}
\begin{center}
\epsfxsize = 10.0cm 
\epsffile{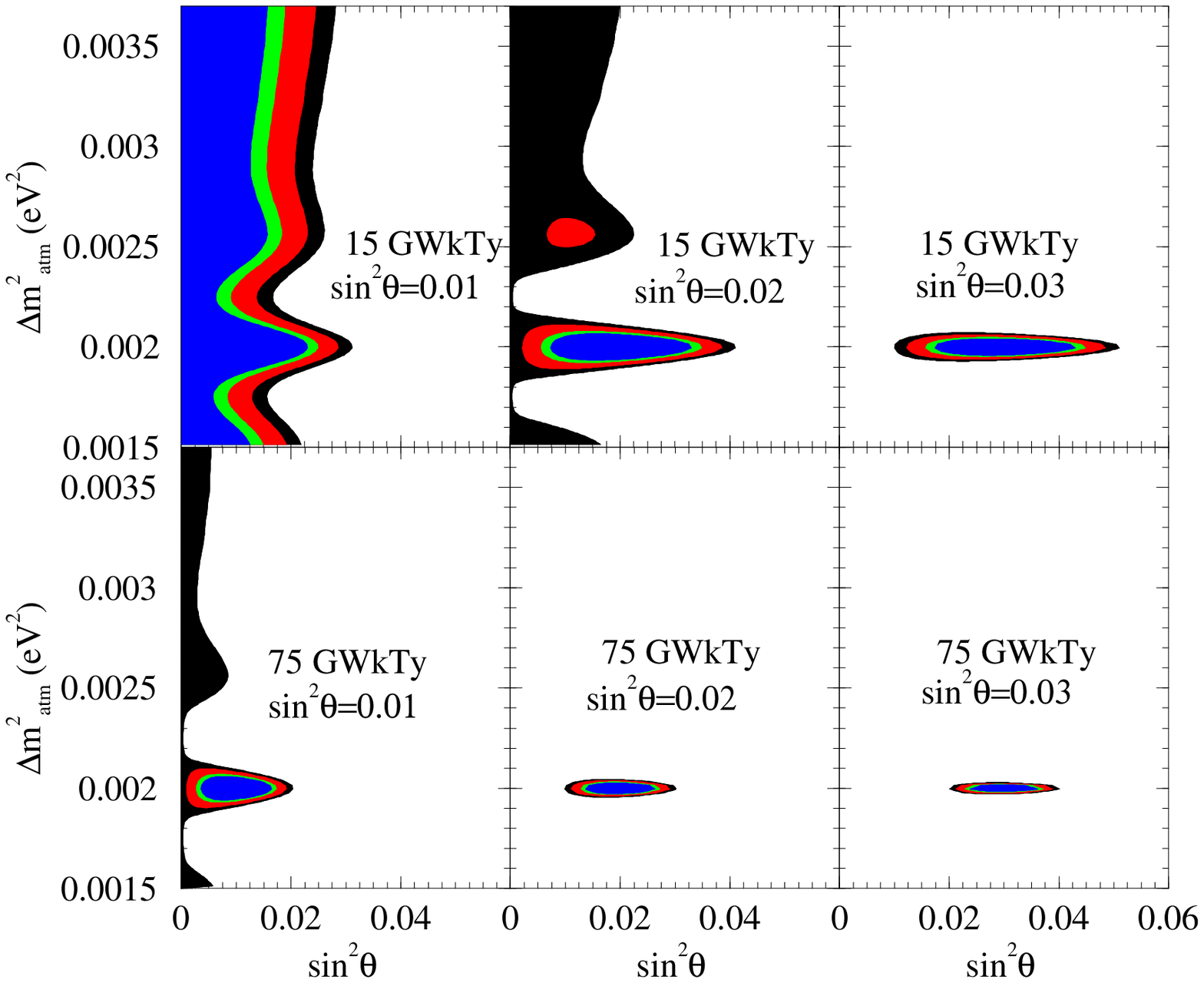}
\leavevmode
\end{center}
\caption{
The same as in Fig.~\ref{fig:d31theta}, 
but assuming that the ``true'' value 
of $\ma=2\times 10^{-3}$ eV$^2$.
}
\label{fig:d31theta2}
\end{figure}

  The experimental setup we 
consider is optimized for extremely precise
measurement of $\theta_{\odot}$, $\ms$, 
and for distinguishing
which hierarchy is realized in the
neutrino mass  
spectrum (as we will see in the next 
Section), provided the high-LMA solution  
is the correct one. Nevertheless, in order
to achieve these goals, a relatively 
large statistics and a sufficiently good energy 
resolution are required.
With these favorable conditions
fulfilled 
and bin width of 0.1 MeV,
one could hope to improve on the 
precision of $\deltaatm$
and $\sch$, provided 
$\sch$ is non-zero 
and is sufficiently large.
{\it We will show that both of these measurements are 
rather independent of the value of 
$\ms$, so that they 
could be performed in this type of an 
experiment even if the low-LMA solution 
is confirmed by the KamLAND
collaboration.}

   In Fig.~\ref{fig:d31theta} we show the 
allowed regions in the 
$\deltaatm -\sch$ plane, 
obtained  for $L=20$ km and two 
different statistics samples: 
the top panels correspond to 15 GWkTy, 
while the lower panels 
are for a 75 GWkTy exposure. 
The results shown in Fig.~\ref{fig:d31theta} 
are for three different ``true'' values
of $\sch$ and for $\ma=2.5\times 10^{-3}$ eV$^2$, 
assuming this to be the ``true'' value. 
Similar results for 
$\ma=2.0\times 10^{-3}$ eV$^2$ and 
the same three values of $\sch$
are presented in 
Fig.~\ref{fig:d31theta2}. 
We take a systematic uncertainty of 2\% 
and consider 0.1 MeV bins spanning the entire 
visible energy spectrum between $(1.0 - 7.2)$ MeV. 
In these figures we keep 
$\ms$ and $\sss$ fixed at 
$1.5\times 10^{-4}$ eV$^2$ and 0.3 respectively.
Since, as we have shown in the previous Section, 
the intermediate baseline experiment itself will restrict the 
allowed values of $\ms$ and $\sss$ to within a few percent of 
their ``true'' value, our results remain practically unchanged 
even if these parameters were allowed 
to vary freely in the analysis. 
As Fig.~\ref{fig:d31theta} demonstrates, 
in the 15 (75) GWkTy case 
the experiment under discussion 
has almost no sensitivity to 
the value of $\ma$ if 
$\sch \ltap 0.02~(0.01)$. 
Note the similarity between these 
limiting values of 
$\sch$ and the limit on $\sch$ which the 
15 (75) GWkTy experiment can provide 
if the ``true'' value of $\sch$ is 0,
given in the previous sub-section.
This implies that below the 
indicated respective 
limiting values of $\sch$, the experiment 
we are discussing
cannot distinguish between a zero and 
a non-zero $\sch$ and thus fails 
to ``see'' the $\ma-$driven oscillations.
The reason for the modest 
sensitivity to $\sch$  
is related to the fact that 
the $L$ is chosen 
for ``our'' experiment 
to be sensitive to the SPMIN 
due to the $\ms-$driven oscillations. 
Correspondingly, 
$L$ is much bigger than 
the $L^\ast$ (cf. \eq{Lstar}),
suitable for the SPMIN associated
with the $\ma-$driven oscillations. 
The lack of accuracy in the
reconstruction of the details of 
a single oscillation period 
(in ${\rm E_{vis}}$)
affects strongly the sensitivity 
to $\sch$ which controls the amplitude
of the oscillations.
However, this should  
not affect the sensitivity to $\ma$,
which determines the oscillation length, 
because of the presence of many oscillation
periods in the relatively large energy
window offered by the spectrum itself. 

   Thus, if $\sch$ is large enough
so that the $\ma-$driven oscillations
can be observed in the experiment
under discussion, $\ma$ can be measured 
with a high precision. 
Indeed, in the case when 
the ``true'' value of $\sch=0.03$, 
$\ma$ can be determined to 
a within few percent accuracy 
at the 99\% C.L. This accuracy 
could be comparable to the 
sensitivity of the JPARC (JHF-SK) 
project to $\ma$~ 
\footnote{The experiment 
we are discussing
has a much larger statistics 
than the planned JPARC (JHF-SK) 
experiment; for the 
precise determination of 
$\ma$ it also requires 
$\sch$ to be sufficiently large.} 
\cite{jhf}.
It is certainly much better than the 
sensitivity to $\ma$  of the $L=1.7$ km 
short baseline experiments 
\cite{Lindner} with the same statistics. 
This is somewhat unexpected 
since the $L=1.7$ km experiment 
is optimized to see the 
$\ma-$driven oscillations.
However, one should keep in mind 
that the $L=1.7$ km baseline experiment 
cannot constrain the solar 
neutrino oscillation parameters. 
The uncertainty in $\ms$ 
\footnote{For this uncertainty the authors of 
\cite{Lindner} have used  the current 
$3\sigma$ uncertainty on $\ms$, found
in the combined solar neutrino and \kl 
data analysis.} 
allows $\ms$ to take on 
values as high as $(2 - 3)\times 10^{-4}$ eV$^2$.
This drastically 
reduces the sensitivity of 
the $L=1.7$ km experiment to $\ma$.
In the experimental 
set-up we are discussing,
the solar parameters are determined to a very 
high degree of accuracy by the 
experiment itself and 
no external input on their 
errors is required. 
Thus, the intermediate baseline experiment 
has a better sensitivity to $\ma$,
provided $\sch$ is sufficiently large, while
the sensitivity to $\sch$ of 
the shorter baseline experiment is better 
for the reasons discussed above, 
{\it viz.} it is optimized to 
the $\ma-$driven SPMIN.

\begin{figure}
\begin{center}
\epsfxsize = 11cm 
\epsffile{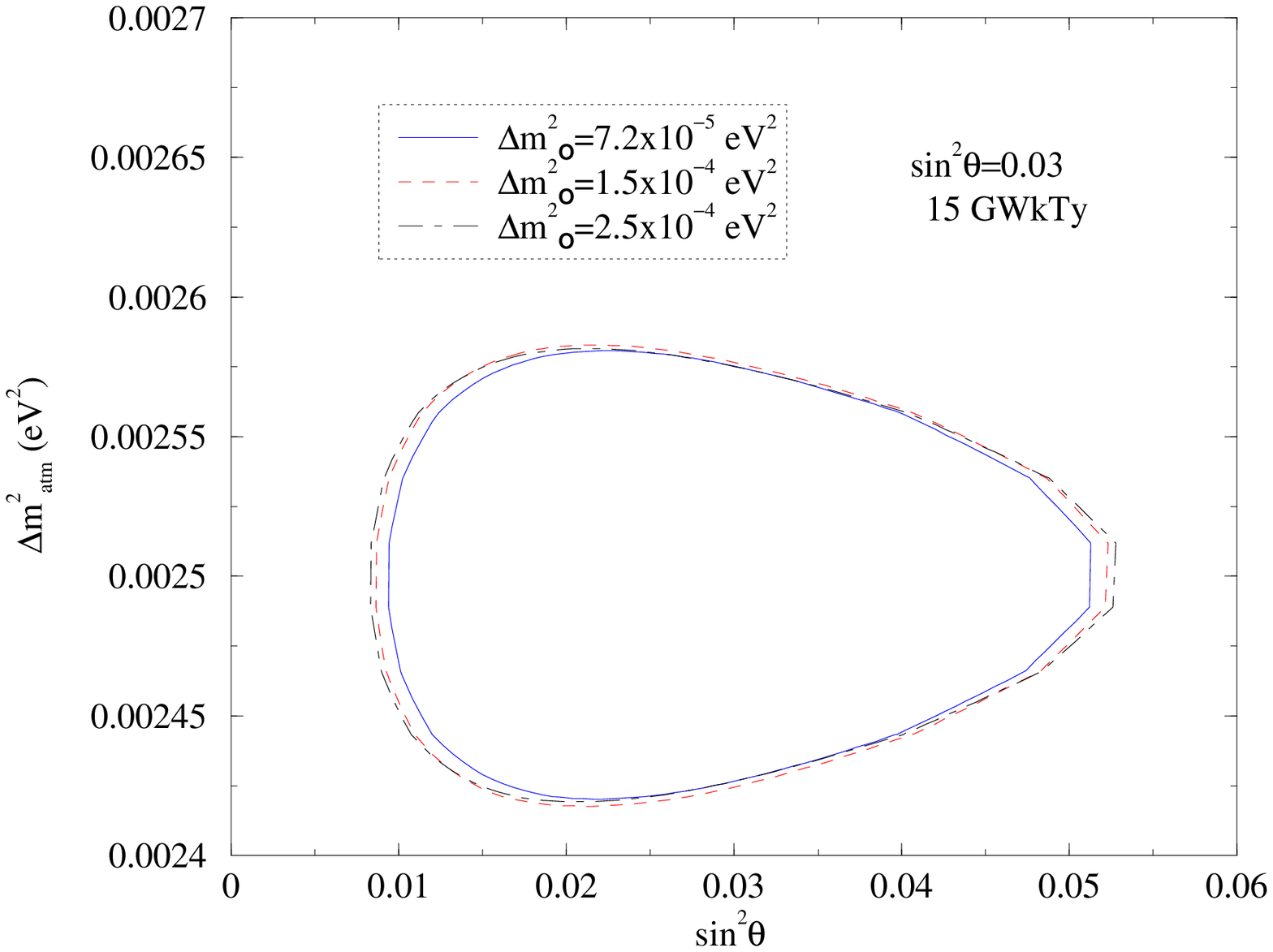}
\leavevmode
\end{center}
\caption{
The impact of the ``true'' value of $\ms$ on the 
99.73\% allowed areas when the ``true'' $\ma=2.5\times 10^{-3}$ eV$^2$ 
and $\sch=0.03$. 
}
\label{fig:d31theta_llma}
\end{figure}

  Finally, we show in Fig.~\ref{fig:d31theta_llma} 
the impact of the ``true'' value 
of $\ms$ on the ability of 
the experiment under consideration
to measure $\ma$ and $\sch$. 
The figure very clearly demonstrates 
that even if the low-LMA solution is confirmed 
as the true solution of the 
solar neutrino problem, the intermediate 
baseline experiment could 
still be used to measure $\ma$ and/or constrain $\sch$.

   We stress that  Figs.~\ref{fig:d31theta}, \ref{fig:d31theta2}  
and \ref{fig:d31theta_llma}
are obtained using \eq{P21sol}, i.e., 
assuming that the neutrino mass 
spectrum conforms to a NH. 
Our conclusions for the sensitivity to 
$\sch$ and $\ma$ would remain the 
same for the case of IH mass spectrum.
For a given ``data'' set, the 
allowed region in the parameter space, 
especially the fitted values of $\ma$, 
depend on the hierarchy assumed 
if the $\ms$ is in the high-LMA region.
The possibility of two types of 
neutrino mass hierarchy 
leads to an ambiguity in the allowed 
values of $\ma$ obtained, 
as we will show in the next sub-section. However, 
even though the assumption of the ``wrong'' hierarchy 
can lead to another (separated) 
allowed region in the parameter space, 
it does not describe the data 
so well as the ``correct'' hierarchy
and is disfavored, in general. This  can be 
used to gain insight into the type of hierarchy the 
neutrino mass spectrum has and can allow 
``our'' intermediate baseline experiment to 
determine the hierarchy. 

   Let us stress that the question of the hierarchy 
becomes relevant if 
$\ms$ lies in the high-LMA zone. 
This will be the subject of our 
discussion in the next subsection.
If the true solution turns out to be 
low-LMA, the last terms in \eq{P21sol} 
and \eq{P32sol} 
would be negligible 
and sensitivity to 
hierarchy would be lost. 
Let us emphasize, however, 
that even in this case it would still be 
possible to improve the existing limits
on $\sch$, and -- if $\sch$ is sufficiently 
large -- to measure $\ma$ with an 
exceptional precision.


\subsection{Normal vs. Inverted Hierarchy}
\label{section:NHvsIH} 


%
\begin{figure}
\begin{center}
\epsfysize = 10cm
\epsfxsize = 17cm \epsffile{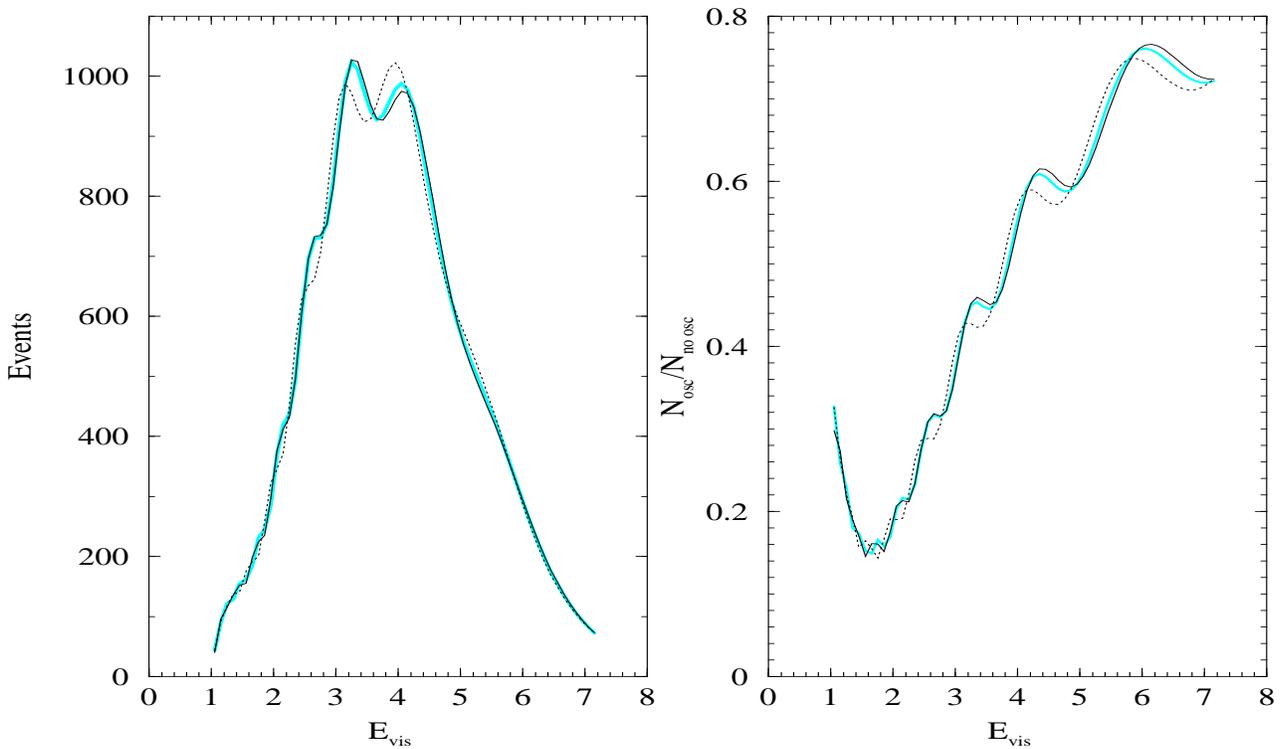}
\leavevmode
\end{center}
\caption{The energy spectrum (in 0.1 MeV bins) of the 
events in the case of 
$\bar{\nu}_e$ oscillations
(left-panel), and 
the ratio of events in the cases of 
oscillations and absence of oscillations 
(right-panel), showing 
the effect of hierarchy for 
a $L=20$ km experiment.
The thick cyan/grey line corresponds to 
the case for NH with $\ma=2.5\times 10^{-3}$ eV$^2$ while the 
dotted and thin solid line correspond to the case of IH with 
$\ma=2.5\times 10^{-3}$ eV$^2$ and $\ma=2.6\times 10^{-3}$ eV$^2$,
respectively. 
The statistics assumed is ${\cal L}=75$ GWkTy, while $\sch=0.03$, 
$\ms=1.5\times 10^{-4}$ 
eV$^2$ and $\sss=0.3$.
}
 \label{fig:spechier}
\end{figure}

   The next question that may be 
answered with the experiment
under discussion is
whether the neutrino mass spectrum is 
with normal hierarchy or with inverted hierarchy.
As pointed out in Section \ref{section:numerical} 
and follows from \eq{interference}, 
the baseline $L=20 \div 30$ km is 
particularly suited for that purpose.
In this subsection we present  
results for $L=20$ km.
We use the highest statistics 
and smallest energy bins and 
consider the entire visible energy 
spectrum with a low energy cut-off
of $E_{th} = 1.0$ MeV.
We assume also that the ``true'' 
value of $\sch$ is non-zero and 
is sufficiently large. 
As it follows from \eq{P21sol} 
and \eq{P32sol}, the predictions 
for the final state $e^+-$spectrum distortions 
in the NH and IH cases differ only if the solar 
neutrino mixing angle 
$\theta_{\odot} \neq \pi/4$. 
Maximal mixing is currently disfavored 
by the solar neutrino and KamLAND data \cite{SolFit1,SolFit2,SolFit3}.
We use the current best-fit value
for the solar neutrino 
mixing angle, corresponding to
$\sss=0.3$. The value of  
$\sss$ will be 
measured with a very high precision 
in the reactor experiment  
under discussion itself.

   In Fig.~\ref{fig:spechier} we show 
the visible energy spectrum expected 
for both the NH and IH cases. The left-hand 
panel shows the number of events, 
while the right-hand panel shows the 
ratio of the number of events in the case 
of $\bar{\nu}_e$ oscillation  
(``oscillation events'')
to the number of events in 
the absence of oscillations 
(``no oscillation events''). 
The thick cyan/grey line is for 
the case of $\ma=2.5\times 10^{-3}$ eV$^2$ 
and NH, the dotted line 
corresponds to $\ma=2.5\times 10^{-3}$ eV$^2$ 
and IH, while the thin solid line 
is for $\ma=2.6\times 10^{-3}$ eV$^2$ and IH.
Thus, for the same value of $\ma$, 
the $e^+-$spectrum deformations 
expected for NH and IH are different 
owing to the last terms in 
\eq{P21sol} and \eq{P32sol}. 
One might expect  
on the basis of Eqs. (\ref{P21sol}) and
(\ref{P32sol})
that the IH $e^+-$spectrum would 
fit a NH $e^+-$spectrum with a value of 
$\ma$ that is larger than the 
``true'' value by a range which would 
be of the order of $\ms$. 
If the experiment has a sensitivity 
to $\ma$ which is better than, 
or is at least of the same order as $\ms$,  
one might expect two non-degenerate 
solutions in the $\ma-\sch$
parameter space for the same data 
set: one for NH and another for IH. 

  However, {\it we would like to stress 
that even though the 
IH $e^+-$ spectrum could approximately 
reproduce the NH $e^+-$spectrum, 
it cannot exactly reproduce the latter. 
Hence, if the NH is the true 
hierarchy, the case of IH mass spectrum 
would always be disfavored by the data
from ``our'' experiment 
compared to the NH spectrum}.
We have seen in the 
previous Section that the experimental 
set-up we are discussing could have a 
very high sensitivity to $\ma$,
and therefore we can expect the large 
statistics intermediate baseline 
reactor experiments to determine the 
type of neutrino mass hierarchy, 
for sufficiently large $\sch$ at least, 
if the high-LMA solution holds.

%
\begin{figure}
\begin{center}
\epsfxsize = 10cm \epsffile{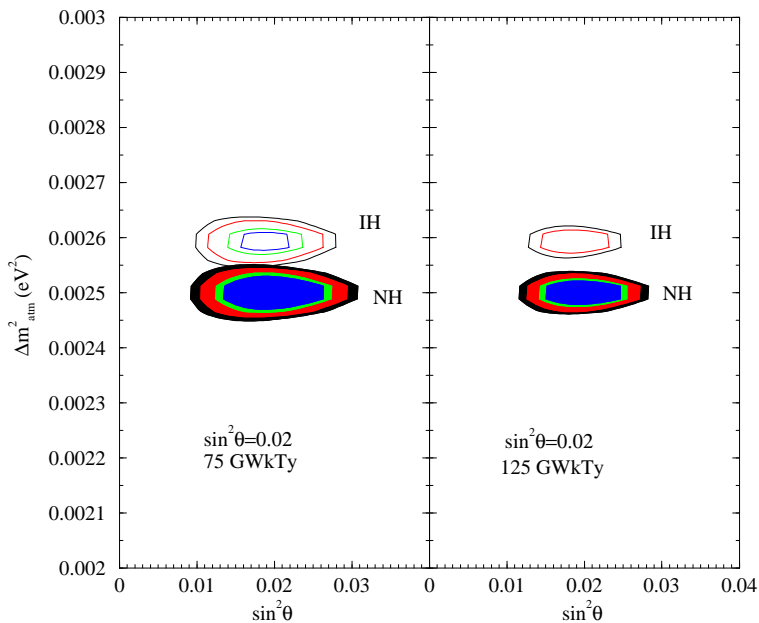}
\leavevmode
\end{center}
\caption{The same as in Fig.~\ref{fig:d31theta}, 
but probing the type of the neutrino mass hierarchy.
The ``data'' sets simulated correspond to NH.
The filled contours show the C.L. 
allowed regions obtained by 
fitting for the parameters assuming NH. 
The empty contours show the 90\%, 95\%, 99\% and 99.73\% 
solution regions, 
obtained by fitting the NH ``data'' 
with an IH theory. The contours for the 
IH are obtained with respect to the 
$\chi^2$-minima found in the case of the NH.
The statistics assumed in the left-hand 
panel is ${\cal L}=75$ GWkTy, 
while that in the right-hand 
panel is ${\cal L}=125$ GWkTy. 
We note that while for the 
${\cal L}=75$ GWkTy case the ``wrong'' 
hierarchy is allowed even at the 90\% C.L., 
with the ${\cal L}=125$ GWkTy statistics 
the experiment can rule out the IH 
at least at the 95\% C.L. 
}
\label{fig:conthier}
\end{figure}

  In Fig.~\ref{fig:conthier} we explicitly 
show the allowed regions obtained 
using a ``data'' set generated in the NH 
case and fitted by both the NH and 
IH expressions for the $\bar{\nu}_e$ 
survival probability.
The results shown are obtained using 
the 75 GWkTy statistics and 
a higher 125 GWkTy statistics. 
The ``true'' value of $\ma$ assumed is 
$2.5\times 10^{-3}$ eV$^2$, while 
``true'' $\sch=0.02$. For the 75 GWkTy 
case, the two regions overlap at 
the $3\sigma$ level, while for 
125 GWkTy they are completely 
non-degenerate. Note also that 
for the 125 GWkTy case, the ``data'' 
generated for NH spectrum can ``rule out'' 
the IH spectrum, at least at the 95\% C.L.

%
\begin{figure}
\begin{center}
\epsfysize = 10cm
\epsfxsize = 16.5cm \epsffile{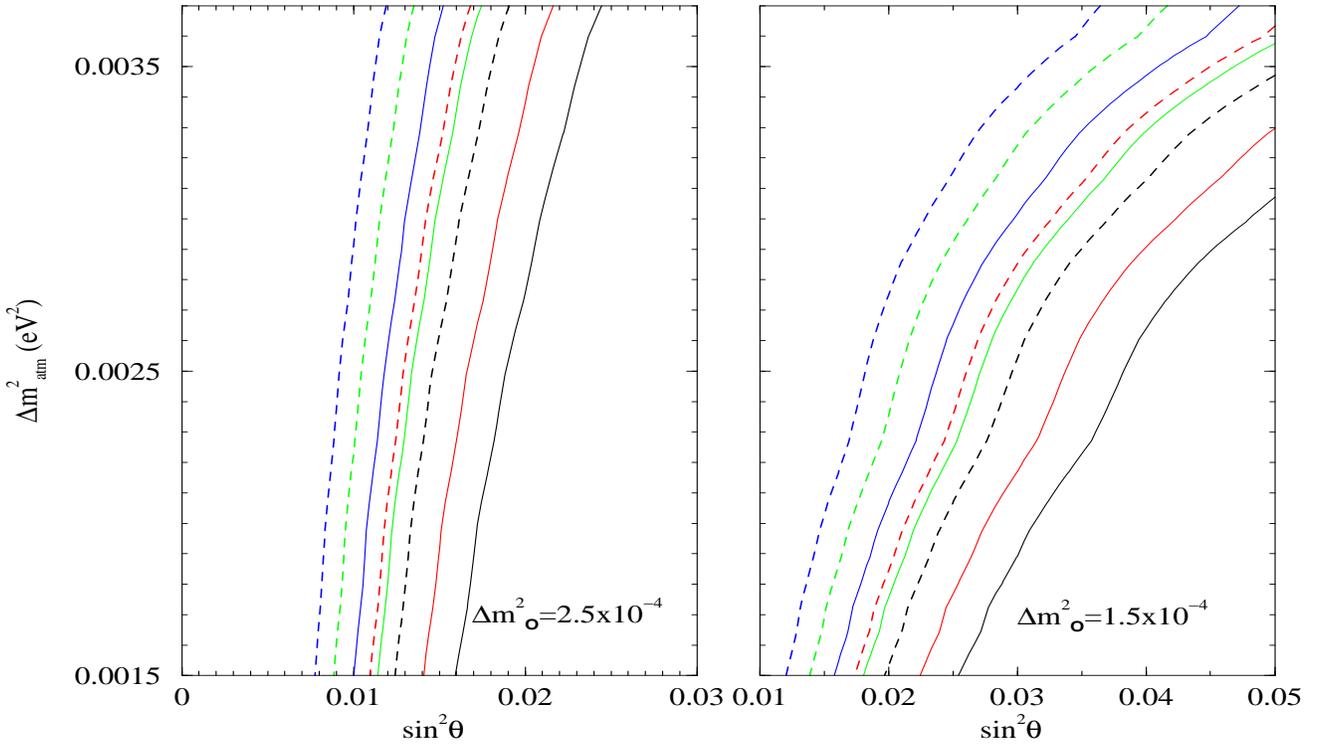}
\leavevmode
\end{center}
\caption{Contours of constant 99.73\%(black), 
99\%(red), 95\%(green) 
and 90\%(blue) C.L.,
showing the
ability
of the $L=20$ km baseline 
experiment to determine the correct 
hierarchy at each ``true'' point in 
the $\ma-\sch$ plane. The solid
lines (dashed lines) correspond to ${\cal L}=75$ GWkTy 
(${\cal L}=125$ GWkTy).
The results shown are for
$\sin^2\theta_{\odot} = 0.30$, and
for $\ms=1.5 \times 10^{-4}$ eV$^2$
(right panel)
 and $\ms=2.5 \times 10^{-4}$ eV$^2$
(left panel). For the values of
$\ma$ and $\sch$, corresponding
to points on the plane, located
to the right of a given C.L. line, 
the experiment can rule out the
``wrong'' hierarchy at that C.L.
See the text for further details. 
}
 \label{fig:sensehier}
\end{figure}
\begin{figure}
\begin{center}
\epsfxsize = 12cm \epsffile{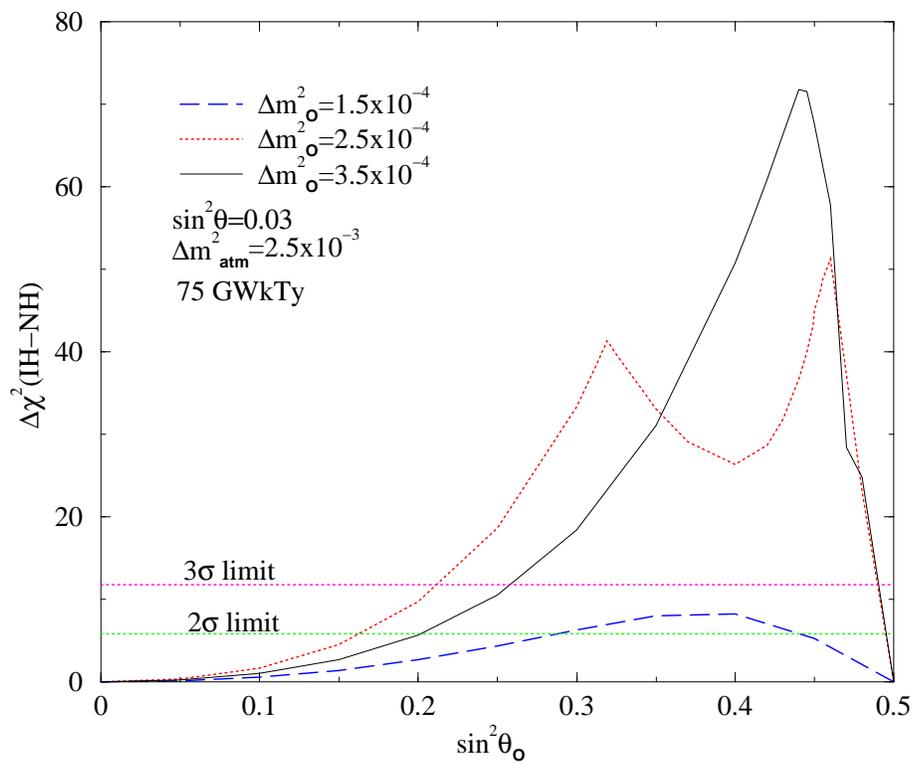}
\leavevmode
\end{center}
\caption{The difference between the 
minimal values of $\chi^2$, obtained by fitting a 
NH ``data'' set with  NH and IH spectrum, as a function of 
$\sss$. The results shown are
for $\ms=1.5\times 10^{-3};~2.5\times 10^{-3};
~3.5\times 10^{-3}$ eV$^2$.
The ``true'' values of $\ma=2.5\times 10^{-3}$ eV$^2$ 
and of $\sch=0.03$, while the statistics 
corresponds to ${\cal L}=75$ GWkTy.
}
 \label{fig:hierth12}
\end{figure}

  In Fig.~\ref{fig:sensehier} we show 
the dependence of the hierarchy 
sensitivity of the $L=20$ km reactor 
experiment on the ``true'' values of $\ma$, $\sch$ and $\ms$.
For each point in the $\ma-\sch$ parameter space and a given 
$\ms$ and $\sss$, we simulate the 
observed $e^+-$energy spectrum assuming NH to be true.
We then fit this ``observed'' spectrum with both the NH and IH through a 
$\chi^2$ analysis, allowing the parameters $\ma$ and $\sch$ to vary 
freely around their ``true'' values.
The IH spectrum obviously does not fit the ``data'' as well as the 
NH and hence is disfavored. 
To quantify this sensitivity to hierarchy we 
calculate the difference ($\chi^2_{diff}$) between the 
$\chi^2$ for the NH and IH for each point in the parameter 
space and use 
the definition $\chi^2_{diff} <\Delta \chi^2$ 
to find whether IH can be ruled out 
at a given C.L. The $\Delta \chi^2$ is determined by the C.L. 
considered and we take the 2-parameter definition for it. 
We show the lines of the C.L. contours in the 
$\ma-\sch$ plane for 2 different values of $\ms$ and $\sss=0.3$. 
All ``true'' values 
of the parameters located to the right of a 
given line, could allow the experiment to
disfavor the IH spectrum 
at the corresponding C.L.
The solid lines give the sensitivities 
for 75 GWkTy exposure, while the dashed 
lines are for a 125 GWkTy 
data set. The
sensitivity to hierarchy 
is maximal for the smallest 
allowed values of $\ma$
and for these cases it is 
possible to rule out IH even for 
smaller values of $\sch$.
It becomes increasingly more
difficult to distinguish 
between the two types 
of neutrino mass hierarchy
as $\ma$ increases, as the figure indicates.  
The sensitivity of 
the experiment to hierarchy depends 
critically on the ``true'' value of 
the solar parameters, as is evident 
from \eq{P21sol} and \eq{P32sol}.
In particular, higher values of $\ms$ 
are better suited for 
hierarchy determination, as can be
seen explicitly comparing the two panels 
in Fig.~\ref{fig:sensehier}. 

   The possibility of determining the
neutrino mass hierarchy is very sensitive
to the value of $\sin^2\theta_{\odot}$.
As we have explained above, when we 
generate the NH ``data'' and fit them
with the IH theory, we allow 
$\ma$ to vary and take larger values
than $\ma$ used to generate the ``data''.
Let us denote the best-fit
$\ma$ in the IH case as 
$\Delta m'^2_{{\rm atm}}$:
\begin{equation}
\Delta m'^2_{{\rm atm}} = \ma + \Delta m^2,
\end{equation}
%
\noindent where $\Delta m^2 > 0$.
The difference between the IH and NH 
$\bar{\nu}_e$ survival probabilities
at each $E$ is given by:
\bea
{P_{IH}({\bar \nu_e}\to{\bar \nu_e})}-
{P_{NH}({\bar \nu_e}\to{\bar \nu_e})}
=&& 4\sch \cos^2\theta \bigg[\sin(\frac{\ma L}{2E} + \frac{\Delta m^2L}{4E})
\sin(-\frac{\Delta m^2L}{4E})
\nonumber\\
&&+\cos^2\theta_\odot \sin(\frac{\ma L}{2E} + \frac{\Delta m^2L}{2E} 
-\frac{\Delta m^2_\odot L}{4E})
\sin(\frac{\Delta m^2_\odot L}{4E})
\nonumber\\
&&
-\sin^2\theta_\odot \sin(\frac{\ma L}{2E}-\frac{\Delta m^2_\odot L}{4E})
\sin(\frac{\Delta m^2_\odot L}{4E})\bigg]
\label{sth12}
\eea
%
\noindent 
When, e.g., $\cos2\theta_{\odot} = 1$,
we have $P_{IH}-P_{NH} = 0$
for $\Delta m^2 = \Delta m^2_{\odot}$.
For $\cos2\theta_{\odot} = 0$ (maximal mixing),
the difference between the two probabilities is
0 for $\Delta m^2 = 0$. For the realistic case
of $\cos2\theta_{\odot} \cong (0.10 - 0.40)$,
it is impossible to have 
$P_{{\rm IH}} - P_{{\rm NH}} = 0$
for every value of $E$ from 
the interval of interest. 
This is illustrated in Fig.~\ref{fig:hierth12}, 
where we show the difference
between the minimal values of
$\chi^2$ for the ``wrong'' IH
and the ``right'' NH spectra
as a function of $\sss$ for 
``true'' $\ms=1.5\times 10^{-4};~
2.5\times 10^{-4};~3.5\times 10^{-4}$ eV$^2$, 
$\ma=2.5\times 10^{-3}$ eV$^2$ and $\sch=0.03$.
As Fig.~\ref{fig:hierth12} indicates,
the best sensitivity to 
the type of neutrino
mass hierarchy is achieved in the interval
$\sin^2\theta_{\odot} \cong (0.3 - 0.45)$,
which is favored by the current solar neutrino and 
KamLAND data. 
The rather strong dependence
of the sensitivity to the type of neutrino mass hierarchy
on $\sin^2\theta_{\odot}$ is illustrated 
in Fig. 18, in which we show contours of
constant C.L. in the $\ma - \sin^2\theta$ plane, 
at which one could distinguish between the NH and IH
spectra, for 
$\ms = 3.5\times 10^{-4}$ eV$^2$ and 
$\sin^2\theta_{\odot} = 0.30;~ 0.40$.
\begin{figure}
\begin{center}
\epsfxsize = 10cm \epsffile{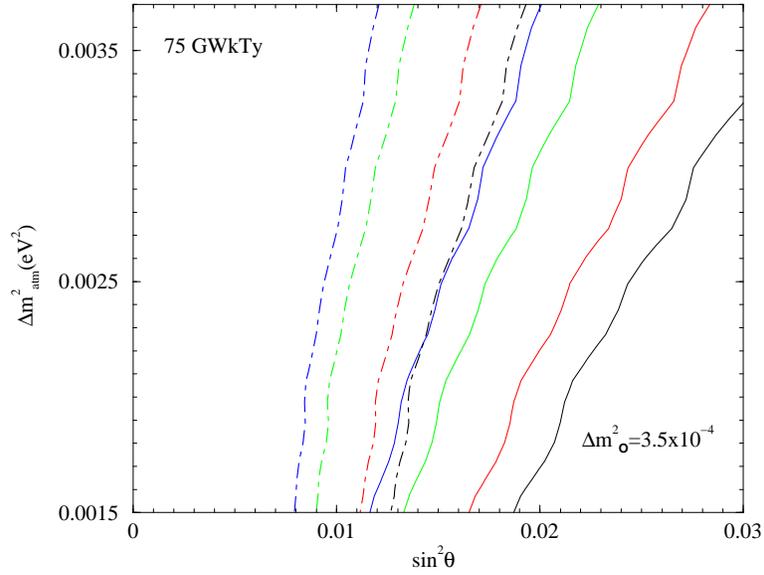}
\leavevmode
\end{center}
\caption{
The same as in Fig. \ref{fig:sensehier}, for
${\cal L}=75$ GWkTy,   
$\ms = 3.5\times 10^{-4}$ eV$^2$ and two
values of $\sin^2\theta_{\odot}$: 
0.30 (solid lines) and 0.40 (dash-dotted lines).
}
\end{figure}

   One final comment is in order.
As Fig.~\ref{fig:conthier} illustrates,
for a given ``data set'' (i.e., ``true'' value 
of $\ma$, etc.)
and sufficient statistics,
the neutrino mass hierarchy ambiguity 
can lead to two non-degenerate solutions for
$\ma$, one for NH and another for IH spectrum.
In certain of these 
cases the intermediate baseline 
reactor experiment might not be able to 
rule out the ``wrong'' hierarchy even at the
90\% C.L. (Fig. \ref{fig:conthier}, left panel).
However, since the NH and IH  give 
two different and non-degenerate
values for $\ma$, in such cases one can use the  
information from some other
very precise experiment, like the proposed
JHF-SK or NuMI off-axis experiments with neutrino 
superbeams  \cite{jhf,Ayres:2002nm},
to determine the correct value of
$\ma$. The latter can be used
together with the data from the reactor
experiment to get information on the 
hierarchy.


\section{Conclusions}
\label{section:conclusions}

%
   In this paper we analyzed the physics potential
of a reactor neutrino experiment 
with a relatively large detector at
a distance $L \sim 20 \div 30$ km. 
This distance has been chosen in order 
to achieve the best sensitivity 
to the solar neutrino oscillation parameters
$\ms$ and $\sin^2\theta_{\odot}$,
assuming that the latter lie in the 
high-LMA solution region.
We considered the case of 
three flavor neutrino mixing
and used in the analysis the exact expression
for the relevant $\bar{\nu}_e$ survival probability. 
The latter depends, in addition to
$\ms$ and $\sin^2\theta_{\odot}$,
on $\deltaatm$, driving the atmospheric 
neutrino oscillations, on $\theta$ -- the 
angle limited by the
CHOOZ and Palo Verde experiments,
and for $\sin^2\theta_{\odot} < 0.50$ -- 
on the type of the neutrino mass spectrum
which can be with normal hierarchy (NH) or
with inverted hierarchy (IH).
The current solar neutrino and KamLAND 
data favor $\sin^2\theta_{\odot} \sim 0.30$.
We discussed strategies 
and the experimental set-up,
which would permit to measure
$\ms$ and $\sin^2\theta_{\odot}$
with a high precision,
get information on (or even measure) 
$\sin^2\theta$, and if 
$\sin^2\theta$ is sufficiently large
($\sin^2\theta \gtap 0.02$)
provide a high precision measurement
of $\deltaatm$ and determine the type of the
neutrino mass hierarchy.
More specifically, we have investigated
the impact that i) the choice of the baseline
$L$, ii) the effect of using a relatively low
$e^+-$ energy cut-off of $E_{th} \sim 1.0$ MeV,
iii) the detector's energy resolution, 
as well as iv) the statistical and systematical
errors, can have on the measurement
of each of the indicated 
neutrino oscillation parameters.
 
  The precision with which $\ms$ and
$\sin^2\theta_{\odot}$ can be determined in the 
experiment under discussion depends crucially
on whether the first $\ms-$driven 
oscillation minimum of the $\bar{\nu}_e$
survival probability falls or not 
in the interval of $e^+-$energies
which can be used for the 
measurements in the experiment.
For $\ms = 1.5 \times 10^{-4}~{\rm eV^2}~
(2.5 \times 10^{-4}~{\rm eV^2})$ and
$L=20$ km, this minimum takes place at
the $e^+-$energy of $E_{vis} = 1.6~(3.2)$ MeV;
for $L=30$ km, it occurs at 
$E_{vis} = 2.8~(5.3)$ MeV.
This implies that if 
$\ms = 1.5 \times 10^{-4}~{\rm eV^2}$,
in order to achieve a high precision in 
the determination of $\ms$ and especially 
of $\sin^2\theta_{\odot}$ in an experiment
with a baseline of $L=20$ km,
a relatively low ``threshold''
energy should be employed,
$E_{th} \sim 1.0$ MeV. If, however,
$\ms \gtap 2.0 \times 10^{-4}~{\rm eV^2}$,
one can use $E_{th} \cong 2.6$ MeV,
as was done in the KamLAND experiment.
In the case of $L=30$ km  
one can use $E_{th} \cong 2.6$ MeV
even for $\ms \gtap 1.5 \times 10^{-4}~{\rm eV^2}$.
If the condition under discussion
is fulfilled,
remarkable precision in the
measurement of $\ms$ and $\sin^2\theta$
can be achieved. For the lower ``threshold''
energy, $E_{th} \sim 1.0$ MeV,
achieving high precision requires
a controllable or negligible background
due to the geophysical neutrinos
in the interval $E_{vis} = (1.0 - 2.6)$ MeV.
If a sufficiently 
accurate modeling of this 
background is achieved and $E_{th} = 1.0$ MeV,
assuming $\ms = 1.5  \times 10^{-4}$ eV$^2$, 
energy bins of $\Delta E = 0.425$ MeV,
$\sin^2\theta_{\odot} = 0.3$,
$2\%$ systematic error
and a total statistics corresponding 
to ${\cal{L}} = 15$ GWkTy,
one could determine 
$\ms$ with up to a $\sim 2\%$,
and $\sin^2 \theta_{\odot}$ 
with a $3 \div 4 \%$ uncertainty at 99\% C.L.
in an experiment with a baseline of $L=20$ km 
(Figs.~\ref{fig:solLEth} - 5). 
This impressive results depend rather mildly 
on the systematic error 
provided the latter does not 
exceed $\sim 4\%$ 
(Fig.~\ref{fig:solSys}).
The reduction of the 
$e^+-$energy bin width leads 
to a negligible change in the precision 
with which $\ms$ and $\sin^2 \theta_{\odot}$
are measured (Fig.~\ref{fig:solBin}).
The precision can be even higher  
in the case of a larger statistics 
(e.g., for ${\cal{L}} = 75$ GWkTy, 
Fig.~\ref{fig:solStat}). 

   The same apparatus could be 
used to extract information on
the other parameters entering into the expression 
for the survival probability, through the
study of sub-leading effects controlled by the (small) mixing
angle $\theta$, for which only upper bounds exist at present.

  The limit on $\sin^2\theta$ can be lowered to 
$\sin^2\theta < 0.021$ at $99.73\%$ C.L. for $L=20$ km,
if the ``threshold'' energy $E_{th} = 1.0$ MeV, 
the systematic error is $2\%$,
the $e^+-$energy bin width is $\Delta E = 0.1$ MeV
and if statistics 
corresponding to ${\cal{L}} = 15$ GWkTy 
is collected. This result depends 
strongly on the choice of $L$: 
the optimal distance for
such a measurement is of a few km, and 
the shorter the distance with respect to
$L=20$ km the stronger this bound would be.  
The magnitude of $e^+-$energy bin width 
and the value of the $e^+-$energy 
effective threshold have a considerable 
impact on the bound as well, while
a further reduction of 
the systematic errors could lead only
to a mild improvement. 
If the statistics is as large as that 
corresponding to a set up 
with ${\cal{L}} = 75$ GWkTy, one could improve
the upper bound to $\sin^2\theta <0.010~(0.0055)$ at 
$99.73\%~(90\%)$ C.L.

  If, on the contrary, a relatively 
large non-vanishing value of
$\sin^2\theta$ is found, 
the experiment could gain 
sensitivity to $\deltaatm$ 
through the detection of the  
sub-leading oscillatory properties 
of the $\bar{\nu}_e$ survival probability.
Not averaging out the $\deltaatm-$driven 
oscillations is crucial for the measurement
of $\deltaatm$. This requires
a relatively high $e^+-$energy resolution, 
permitting a binning in the $e^+-$energy
with a rather small bin width, 
$\Delta E \cong 0.1$ MeV.
Taking $E_{th} = 1.0$ MeV and
assuming $2\%$ systematic error, we find that
a set-up with ${\cal{L}} = 15~(75)$ GWkTy 
could provide a precise determination 
of $\ma$ with a few percent uncertainty
for $\ma \sim 2.5 \times 10^{-3}$ eV$^2$, 
if $\sin^2\theta \sim 0.03~(0.02)$
(Fig.~\ref{fig:d31theta}).
  
   Finally, the distance $L=20$ km 
is the  optimal one 
for distinguishing between the 
NH and IH neutrino mass spectrum.
The relevant $\bar{\nu}_e$ survival probabilities
for the two types of spectrum
differ for $\sin^2\theta_{\odot} <0.5$ 
by a $\sin^2\theta-$suppressed interference term
due to the amplitudes of the
$\ms-$  and $\deltaatm-$ driven 
oscillations. 
For this highly challenging study, 
a very large statistics is necessary: 
we considered ${\cal{L}} = 75$ and $125$ GWkTy. 
The possibility of distinguishing 
between the two types of neutrino mass 
hierarchy was found to depend 
not only on the values of 
$\sin^2\theta$ and $\sin^2\theta_{\odot}$,
but also on the values of $\ms$ and $\ma$.
The dependence on $\ms$ and
$\sin^2\theta_{\odot}$ 
is particularly strong.
For $\ms = 1.5 \times 10^{-4}~{\rm eV^2}$, 
and ${\cal{L}} = 75~(125)$ GWkTy, one could
distinguish the NH from the IH spectrum
at $99.73\%$ C.L. in the region 
of $\ma \ltap 2.5 \times 10^{-3}~{\rm eV^2}$ 
if $\sin^2\theta \gtap 0.038~(0.03)$ 
(Fig.~\ref{fig:sensehier}). 
The type of the neutrino mass spectrum 
can be determined for 
considerably smaller values 
of $\sin^2\theta$
if $\ms$ has a larger value
(Figs. 17 and 18), or
if an even larger statistics
would be accumulated.
For ${\cal{L}} = 75$ GWkTy and
$\ms = 2.5 \times 10^{-4}~{\rm eV^2}$,
for instance, this can be done
at $99.73\%$ C.L. 
for $\sin^2\theta \gtap 0.019$
if $\ma \ltap 2.5 \times 10^{-3}~{\rm eV^2}$
and $\sin^2\theta_{\odot} = 0.30$ 
(Fig.~\ref{fig:sensehier}).
As we have shown, the highest sensitivity
to the type of the neutrino mass hierarchy
is achieved for $\sin^2\theta_{\odot} 
\cong (0.30 - 0.45)$ (Figs. 17 and 18).

 To conclude, the intermediate baseline $L \cong 20$ km
reactor neutrino experiments have in the case of the
high-LMA solution of the solar neutrino problem  
a remarkable potential for
a high precision measurement
of the solar neutrino oscillation parameters
$\ms$ and $\sin^2\theta_{\odot}$,
which can reduce the error 
in the values of the latter 
to a $(2 - 4)\%$. Such an experiment can 
also improve the existing bounds on, or measure,
$\sin^2\theta$, currently limited by the 
CHOOZ and Palo Verde data.
If $\sin^2\theta$ turns out to be relatively large,
$\sin^2\theta \gtap 0.02$,
the same experiment could measure 
$\deltaatm$, driving the 
atmospheric neutrino oscillations,
with an error of a few percent and 
might be able to establish whether the 
neutrino mass spectrum is with normal or 
inverted hierarchy.

\vskip 2mm
\begin{em}
\leftline{\bf Acknowledgements.} 
S.T.P. would like to thank T. Lasserre
for useful discussions and the members
of APC Institute at College de France, Paris, and
the organizers of the Program
on ``Neutrinos: Data, Cosmos and the Planck Scale''
at KITP, Univ. of California at Santa Barbara,
where parts of the work on the present study were done,
for kind hospitality. M.P. would like to thank
R. Shrock and R.G.H. Robertson for useful discussions. 
S.C. acknowledges help from S. Goswami and A. Bandyopadhyay 
in developing the computer code used
for numerical calculations in the present work, 
and would like to thank F. Suekane for useful 
correspondence. This work was supported in part 
by the Italian MIUR and INFN under the programs 
``Fenomenologia delle Interazioni Fondamentali'' 
and ``Fisica Astroparticellare'' (S.T.P. and S.C.),
by the U.S. National Science Foundation under Grant 
No. PHY99-07949 (S.T.P.) and by the US Department of 
Energy under contract DE-FG02-92ER-40704 (M.P.).

\vspace{0.5cm}
\end{em}

%
%



\begin{thebibliography}{50}
%
\parskip = 0pt
\baselineskip = 13pt
%
\medskip
%
\bibitem{sol}
B.~T.~Cleveland {\it et al.},
Astrophys.\ J.\  {\bf 496}, 505 (1998);

J.~N.~Abdurashitov {\it et al.}  [SAGE Collaboration],
arXiv:astro-ph/0204245;

W.~Hampel {\it et al.}  [GALLEX Collaboration],
Phys.\ Lett.\ B {\bf 447}, 127 (1999);

E. Bellotti, Talk at Gran Sasso National Laboratories, 
Italy, May 17, 2002;
T. Kirsten, talk at {\it Neutrino 2002},  
XXth International Conference on
Neutrino Physics and Astrophysics,
Munich, Germany, May 25-30, 2002.{ \it(http://neutrino2002.ph.tum.de/)}

\bibitem{SKsolar}
S.~Fukuda {\it et al.}  [Super-Kamiokande Collaboration],
of  Super-Kamiokande solar neutrino data,''
Phys.\ Rev.\ Lett.\  {\bf 86}, 5656 (2001) [arXiv:hep-ex/0103033];
%
S.~Fukuda {\it et al.}  [Super-Kamiokande Collaboration],
Phys.\ Rev.\ Lett.\  {\bf 86}, 5651 (2001) [arXiv:hep-ex/0103032].


\bibitem{SNO}
Q.~R.~Ahmad {\it et al.}  [SNO Collaboration],
Phys.\ Rev.\ Lett.\  {\bf 87}, 071301 (2001)
[arXiv:nucl-ex/0106015];
%
Q.~R.~Ahmad {\it et al.}  [SNO Collaboration],
Phys.\ Rev.\ Lett.\  {\bf 89}, 011301 (2002)
[arXiv:nucl-ex/0204008].

\bibitem{SKatm}
Y.~Fukuda {\it et al.}  [Super-Kamiokande Collaboration],
Phys.\ Rev.\ Lett.\  {\bf 81}, 1562 (1998) [arXiv:hep-ex/9807003];
                 M.~Shiozawa,
                 talk given at the Int. Conf. on Neutrino Physics and 
                 Astrophysics ``Neutrino'02'', May 25 - 30, 2002, 
                 Munich, Germany.


\bibitem{K2K}
M.~H.~Ahn {\it et al.}  [K2K Collaboration],
Phys.\ Rev.\ Lett.\  {\bf 90}, 041801 (2003)
[arXiv:hep-ex/0212007].



\bibitem{KamLAND}
K.~Eguchi {\it et al.}  [KamLAND Collaboration],
Phys.\ Rev.\ Lett.\  {\bf 90}, 021802 (2003)
[arXiv:hep-ex/0212021].


\bibitem{BPont67} B. Pontecorvo, 
               Zh.\ Eksp.\ Teor.\ Fiz.\ {\bf 53}, 1717 (1967)
               [Sov.\ Phys.\ JETP {\bf 26}, 984 (1968)];
                S. M. Bilenky and B.\ Pontecorvo, Phys. Rep.
                {\bf 41}, 225 (1978).

\bibitem{BiPet87} S.M.\ Bilenky and S.T.\ Petcov,
                {\em Rev.\ Mod.\ Phys.}  {\bf 59} (1987) 671. 

\bibitem{SPSchlad97} 
S.T. Petcov, Lecture Notes in Physics, v. 512
(eds. H. Gausterer and C.B. Lang, Springer, 1998), p. 281.
(hep-ph/9806466).

\bibitem{Choubey:2002nc}
S.~Choubey et al.,
arXiv:hep-ph/0209222;
%
A.~Bandyopadhyay et al., 
Phys.\ Lett.\ B {\bf 540}, 14 (2002)
[arXiv:hep-ph/0204286].



\bibitem{SolFit1} G.~L.~Fogli et al., 
Phys.\ Rev.\ D {\bf 67}, 073002 (2003)
[arXiv:hep-ph/0212127];
%

\bibitem{SolFit2}
A.~Bandyopadhyay et al., 
Phys.\ Lett.\ B {\bf 559}, 121 (2003) [arXiv:hep-ph/0212146];
%
\bibitem{SolFit3}
M.~Maltoni, T.~Schwetz and J.~W.~Valle,
arXiv:hep-ph/0212129;
%
J.~N.~Bahcall, M.~C.~Gonzalez-Garcia and C.~Pena-Garay,
JHEP {\bf 0302}, 009 (2003) [arXiv:hep-ph/0212147];
%
H.~Nunokawa, W.~J.~Teves and R.~Zukanovich Funchal,
arXiv:hep-ph/0212202;
%
P.~Aliani et al., 
arXiv:hep-ph/0212212;
%
P.~C.~de Holanda and A.~Y.~Smirnov,
JCAP {\bf 0302}, 001 (2003) [arXiv:hep-ph/0212270].
%
%

\bibitem{Fogli:2003th} G.~L.~Fogli et al., 
arXiv:hep-ph/0303064.


\bibitem{BGG99} S.\ M.\ Bilenky, C.\ Giunti and W.\ Grimus,
 Prog. Part. Nucl. Phys. {\bf 43} (1999)  1 (hep-ph/9812360).

\bibitem{BPont57} B. Pontecorvo, Zh. Eksp. Teor. Fiz. {\bf 33} (1957) 549,
                and {\bf 34} (1958) 247.

\bibitem{MNS62} Z. Maki, M. Nakagawa and S. Sakata,
Prog. Theor. Phys. 28 (1962) 870.

\bibitem{BHP80} S.M.\ Bilenky \textit{et al.},
              {\em  Phys.\ Lett.}  {\bf B94} (1980) 495.

\bibitem{Doi81} M.~Doi \textit{et al.},
{\em Phys. Lett.}  \textbf{B102} (1981) 323.

\bibitem{Lang86} P. Langacker {\it et al.},
{\it Nucl.Phys.} {\bf B282} (1987) 589.


\bibitem{CHOOZ}
M.~Apollonio {\it et al.}  [CHOOZ Collaboration],
Phys.\ Lett.\ B{\bf 466} (1999) 415 (hep-ex/9907037).
%

\bibitem{PaloV} F. Boehm, J. Busenitz et al., Phys.\ Rev.\ Lett.\  {\bf 84}
(2000) 3764 and Phys. Rev. D{\bf 62} (2000) 072002.


\bibitem{BNPChooz}
S.~M.~Bilenky, D.~Nicolo and S.~T.~Petcov,
Phys.\ Lett.\ B {\bf 538}, 77 (2002) [arXiv:hep-ph/0112216].


\bibitem{ADE80} A. De Rujula \textit{et al.}, 
{\em Nucl. Phys.} {\bf B168} (1980) 54.



\bibitem{solvsreactor}
S.~M.~Bilenky et al., 
Phys.\ Lett.\ B {\bf 356}, 273 (1995) [arXiv:hep-ph/9504405];

\bibitem{MINOS} D. Michael (MINOS Collaboration),
Talk at the Int. Conf. on Neutrino Physics and 
Astrophysics ``Neutrino'02'', May 25 - 30, 2002, Munich, Germany.
                
\bibitem{MSpironu02} M. Spiro, Summary talk at the
Int. Conf. on Neutrino Physics and Astrophysics ``Neutrino'02'',
May 25 - 30, 2002, Munich, Germany.


\bibitem{KlStudies}
A.~Bandyopadhyay et al., 
arXiv:hep-ph/0211266;
%
V.~D.~Barger, D.~Marfatia and B.~P.~Wood,
Phys.\ Lett.\ B {\bf 498}, 53 (2001)
[arXiv:hep-ph/0011251];
%
H.~Murayama and A.~Pierce,
Phys.\ Rev.\ D {\bf 65}, 013012 (2002)
[arXiv:hep-ph/0012075].


\bibitem{Ch}
A.~Bandyopadhyay, S.~Choubey and S.~Goswami,
arXiv:hep-ph/0302243.

\bibitem{SPMPiai01}
S.~T.~Petcov and M.~Piai,
Phys.\ Lett.\ B {\bf 533}, 94 (2002)
[arXiv:hep-ph/0112074].

\bibitem{HLMA}
S.~Schonert, T.~Lasserre and L.~Oberauer,
Astropart.\ Phys.\  {\bf 18}, 565 (2003)
[arXiv:hep-ex/0203013].


\bibitem{BPP1}  
S.M. Bilenky \textit{et al.}, {\em Phys.\ Lett.} 
{\bf B465} (1999) 193;
S.M.\ Bilenky, S. Pascoli  and S.T.\ Petcov,
Phys. Rev. D{\bf 64} (2001) 053010; 
S. Pascoli and  S.T.\ Petcov, 
Phys. Lett. B{\bf 544} (2002) 239; 
S. Pascoli, S.T.\ Petcov and W. Rodejohann,
Phys. Lett. B{\bf 558} (2003) 141.


\bibitem{AMMS99} M. Freund et al., 
Nucl. Phys. {\bf B578}  (2000) 27; see also, e.g.,
A. De Rujula, M.B. Gavela and P. Hernandez,
Nucl. Phys. {\bf B547}, 21 (1999), and
V. Barger et al.,
Phys. Rev. {\bf D62}, 013004 (2000).
%

%





\bibitem{HLM} V. Barger, D. Marfatia and K. Whisnant, hep-ph/0210428;
P. Huber, M. Lindner and W. Winter, hep-ph/0211300.



\bibitem{SChReactP}
A.~Bandyopadhyay, S.~Choubey, S.~Goswami and K.~Kar,
Phys.\ Rev.\ D {\bf 65}, 073031 (2002)
[arXiv:hep-ph/0110307].

\bibitem{BGKP96} S.M. Bilenky et al.,
Phys. Rev. D \textbf{54} (1996) 4432 (hep-ph/9604364).


\bibitem{Kopeikin:2001qv}
V.~I.~Kopeikin, L.~A.~Mikaelyan and V.~V.~Sinev,
Phys.\ Atom.\ Nucl.\  {\bf 64}, 849 (2001) [Yad.\ Fiz.\  {\bf
64}, 914 (2001)] (hep-ph/0110290).




\bibitem{Fiorentini:2003ww}
G.~Fiorentini et al., 
Phys.\ Lett.\ B {\bf 558}, 15 (2003) [arXiv:hep-ph/0301042].


\bibitem{BGV}
C.~Bemporad, G.~Gratta and P.~Vogel,
Rev.\ Mod.\ Phys.\  {\bf 74}, 297 (2002)
[arXiv:hep-ph/0107277].

\bibitem{cc}
M.~C.~Gonzalez-Garcia and C.~Pena-Garay,
Phys.\ Lett.\ B {\bf 527}, 199 (2002) [arXiv:hep-ph/0111432].

\bibitem{Lindner}
P.~Huber et al., 
arXiv:hep-ph/0303232.

\bibitem{mina}
H.~Minakata et al., 
arXiv:hep-ph/0211111.

\bibitem{jhf}
Y.~Itow {\it et al.},
arXiv:hep-ex/0106019
%
\bibitem{Ayres:2002nm}
D.~Ayres {\it et al.},
arXiv:hep-ex/0210005.








\end{thebibliography}
\end{document}